\documentclass[aps,prl,preprint,nopacs,superscriptaddress]{revtex4}
\usepackage{graphicx}
\usepackage{verbatim}
\usepackage{relsize}
\usepackage{mathrsfs}
\usepackage{setspace}
\usepackage{color}
\usepackage{ulem}
\usepackage{amsmath,amsfonts,amssymb}
\usepackage{graphicx}

\definecolor{mygray}{gray}{0.6}
\usepackage{bbding}
\usepackage{pifont}
\usepackage{wasysym}
\usepackage{amssymb}

\def \beq {\begin{equation}}
\def \eeq {\end{equation}}
\begin{document}

\title{Layer Hall effect in a 2D topological Axion antiferromagnet\\}

\author{\footnotesize Anyuan Gao}\affiliation{\setstretch{1.2}\footnotesize Department of Chemistry and Chemical Biology, Harvard University, MA 02138, USA}
\author{\footnotesize Yu-Fei Liu}\affiliation{\setstretch{1.2}\footnotesize Department of Chemistry and Chemical Biology, Harvard University, MA 02138, USA}

\author{\footnotesize Chaowei Hu}
\affiliation{\setstretch{1.2}\footnotesize Department of Physics and Astronomy and California NanoSystems Institute, University of California, Los Angeles, Los Angeles, CA 90095, USA.}

\author{\footnotesize Jian-Xiang Qiu}\affiliation{\setstretch{1.2}\footnotesize Department of Chemistry and Chemical Biology, Harvard University, MA 02138, USA}

\author{\footnotesize Christian Tzschaschel}\affiliation{\setstretch{1.2}\footnotesize Department of Chemistry and Chemical Biology, Harvard University, MA 02138, USA}

\author{\footnotesize Barun Ghosh}\affiliation{\setstretch{1.2}\footnotesize Department of Physics, Indian Institute of Technology, Kanpur, India}\affiliation{\setstretch{1.2}\footnotesize Department of Physics, Northeastern University, Boston, MA 02115, USA}

\author{\footnotesize Sheng-Chin Ho}\affiliation{\setstretch{1.2}\footnotesize Department of Chemistry and Chemical Biology, Harvard University, MA 02138, USA}

\author{\footnotesize Damien B\'erub\'e}\affiliation{\setstretch{1.2}\footnotesize Department of Chemistry and Chemical Biology, Harvard University, MA 02138, USA}

\author{\footnotesize Rui Chen}
\affiliation{\setstretch{1.2}\footnotesize Shenzhen Institute for Quantum Science and Engineering and Department of Physics, Southern University of Science and Technology (SUSTech), Shenzhen 518055, China}
\author{\footnotesize Haipeng Sun}
\affiliation{\setstretch{1.2}\footnotesize Shenzhen Institute for Quantum Science and Engineering and Department of Physics, Southern University of Science and Technology (SUSTech), Shenzhen 518055, China}

\author{\footnotesize Zhaowei Zhang}\affiliation{\setstretch{1.2}\footnotesize Division of Physics and Applied Physics, School of Physical and Mathematical Sciences, Nanyang Technological University, Singapore 637371, Singapore}

\author{\footnotesize Xin-Yue Zhang}\affiliation{\setstretch{1.2}\footnotesize Department of Physics, Boston College, Chestnut Hill, MA, USA}
\author{\footnotesize Yu-Xuan Wang}\affiliation{\setstretch{1.2}\footnotesize Department of Physics, Boston College, Chestnut Hill, MA, USA}

\author{\footnotesize Naizhou Wang}\affiliation{\setstretch{1.2}\footnotesize Division of Physics and Applied Physics, School of Physical and Mathematical Sciences, Nanyang Technological University, Singapore 637371, Singapore}

\author{\footnotesize Zumeng Huang}\affiliation{\setstretch{1.2}\footnotesize Division of Physics and Applied Physics, School of Physical and Mathematical Sciences, Nanyang Technological University, Singapore 637371, Singapore}

\author{\footnotesize Claudia Felser}\affiliation{\setstretch{1.2}\footnotesize Max Planck  Institute for Chemical Physics of Solids, N{\"o}thnitzer Stra{\ss}e 40, 01187 Dresden, Germany}

\author{\footnotesize Amit Agarwal}
\affiliation{\setstretch{1.2}\footnotesize Department of Physics, Indian Institute of Technology, Kanpur, India}
\author{\footnotesize Thomas Ding}\affiliation{\setstretch{1.2}\footnotesize Department of Physics, Boston College, Chestnut Hill, MA, USA}
\author{\footnotesize Hung-Ju Tien}\affiliation{\setstretch{1.2}\footnotesize Department of Physics, National Cheng Kung University, Tainan 701, Taiwan}

\author{\footnotesize Austin Akey}\affiliation{\setstretch{1.2}\footnotesize Center for Nanoscale Systems, Harvard University,Cambridge, MA 02138, USA}
\author{\footnotesize Jules Gardener}\affiliation{\setstretch{1.2}\footnotesize Center for Nanoscale Systems, Harvard University,Cambridge, MA 02138, USA}
\author{\footnotesize Bahadur Singh}\affiliation{\setstretch{1.2}\footnotesize Department of Condensed Matter Physics and Materials Science, Tata Institute of Fundamental Research, Colaba, Mumbai 400005, India}

\author{\footnotesize Kenji Watanabe}\affiliation{\setstretch{1.2}\footnotesize Research Center for Functional Materials, National Institute for Materials Science, 1-1 Namiki, Tsukuba 305-0044, Japan}
\author{\footnotesize Takashi Taniguchi}\affiliation{\setstretch{1.2}\footnotesize International Center for Materials Nanoarchitectonics, National Institute for Materials Science,  1-1 Namiki, Tsukuba 305-0044, Japan}

\author{\footnotesize Kenneth S. Burch}\affiliation{\setstretch{1.2}\footnotesize Department of Physics, Boston College, Chestnut Hill, MA, USA}
\author{\footnotesize David C. Bell}\affiliation{\setstretch{1.2}\footnotesize Center for Nanoscale Systems, Harvard University,Cambridge, MA 02138, USA}
\affiliation{\setstretch{1.2}\footnotesize Harvard John A. Paulson School of Engineering and Applied Sciences,Harvard University, Cambridge, MA 02138, USA}

\author{\footnotesize Brian B. Zhou}\affiliation{\setstretch{1.2}\footnotesize Department of Physics, Boston College, Chestnut Hill, MA, USA}

\author{\footnotesize Weibo Gao}\affiliation{\setstretch{1.2}\footnotesize Division of Physics and Applied Physics, School of Physical and Mathematical Sciences, Nanyang Technological University, Singapore 637371, Singapore}

\author{\footnotesize Hai-Zhou Lu}
\affiliation{\setstretch{1.2}\footnotesize Shenzhen Institute for Quantum Science and Engineering and Department of Physics, Southern University of Science and Technology (SUSTech), Shenzhen 518055, China}

\author{\footnotesize Arun Bansil}
\affiliation{\setstretch{1.2}\footnotesize Department of Physics, Northeastern University, Boston, MA 02115, USA}
\author{\footnotesize Hsin Lin}
\affiliation{\setstretch{1.2}\footnotesize Institute of Physics, Academia Sinica, Taipei 11529, Taiwan}

\author{\footnotesize Tay-Rong Chang}
\affiliation{\setstretch{1.2}\footnotesize Department of Physics, National Cheng Kung University, Tainan 701, Taiwan}\affiliation{\setstretch{1.2}\footnotesize Center for Quantum Frontiers of Research and Technology (QFort), Tainan 701, Taiwan}\affiliation{\setstretch{1.2}\footnotesize Physics Division, National Center for Theoretical Sciences, National Taiwan University, Taipei, Taiwan}

\author{\footnotesize Liang Fu}\affiliation{\setstretch{1.2}\footnotesize Department of Physics, Massachusetts Institute of Technology, Cambridge, MA 02139, USA}

\author{\footnotesize Qiong Ma}\affiliation{\setstretch{1.2}\footnotesize Department of Physics, Boston College, Chestnut Hill, MA, USA}
\author{\footnotesize Ni Ni$^{*}$}\affiliation{\setstretch{1.2}\footnotesize Department of Physics and Astronomy and California NanoSystems Institute, University of California, Los Angeles, Los Angeles, CA 90095, USA.}

\author{\footnotesize Su-Yang Xu\footnote{Corresponding authors (emails): suyangxu@fas.harvard.edu and nini@physics.ucla.edu }}\affiliation{\setstretch{1.2}\footnotesize Department of Chemistry and Chemical Biology, Harvard University, MA 02138, USA}

\maketitle

\textbf{While ferromagnets have been known and exploited for millennia, antiferromagnets (AFMs) were only discovered in the 1930s \cite{neel1972magnetism}. The elusive nature indicates AFMs' unique properties: At large scale, due to the absence of global magnetization, AFMs may appear to behave like any non-magnetic material; However, such a seemingly mundane macroscopic magnetic property is highly nontrivial at microscopic level, where opposite spin alignment within the AFM unit cell forms a rich internal structure. In topological AFMs, such an internal structure leads to a new possibility, where topology and Berry phase can acquire distinct spatial textures \cite{li2013coupling, gao2014field}. Here, we study this exciting possibility in an AFM Axion insulator, even-layered MnBi$_2$Te$_4$ flakes, where spatial degrees of freedom correspond to different layers. Remarkably, we report the observation of a new type of Hall effect, the layer Hall effect, where electrons from the top and bottom layers spontaneously deflect in opposite directions. Specifically, under no net electric field, even-layered MnBi$_2$Te$_4$ shows no anomalous Hall effect (AHE); However, applying an electric field isolates the response from one layer and leads to the surprising emergence of a large layer-polarized AHE ($\sim50\% \frac{e^2}{h}$). Such a layer Hall effect uncovers a highly rare layer-locked Berry curvature, which serves as a unique character of the space-time $\mathcal{PT}$-symmetric AFM topological insulator state. Moreover, we found that the layer-locked Berry curvature can be manipulated by the Axion field, $\mathbf{E}\cdot\mathbf{B}$, which drives the system between the opposite AFM states. Our results achieve previously unavailable pathways to detect and manipulate the rich internal spatial structure of fully-compensated topological AFMs \cite{chen2014anomalous, nakatsuji2015large, nayak2016large, mong2010antiferromagnetic, vsmejkal2018topological, tokura2019magnetic, mong2019magnetic, xu2020high, cheong2020seeing}. The layer-locked Berry curvature represents a first step towards spatial engineering of Berry phase, such as through layer-specific moir\'e potential.\color{black}}

\vspace{0.5cm}

Ever since navigation was achieved by a piece of lodestone, magnetism has played a central role in fundamental research and technology developments. As we enter the quantum era, an important frontier of modern condensed matter is to search for quantum magnets, where electronic correlations, symmetry breakings, Berry phase, etc. interact with magnetism, leading to exotic phenomena that don't exist in conventional magnetic materials. Under such a scope, antiferromagnetism is an elementary yet fascinating type of magnetic order. Antiferromagnets (AFMs) are internally magnetic, but the zero net magnetization makes the magnetism externally invisible. Although the rich internal magnetic structure does not manifest as a global magnetization, it can profoundly affect many other macroscopic properties, leading to novel physics: In strongly-correlated electronic systems, the internal anti-parallel spin structure promotes virtual hopping \cite{lee2006doping}, making AFM a favorable ground state in most un-doped Mott insulators. In some multiferroics, the internal AFM spin structures can break spatial symmetries, leading to finite magnetoelectric coupling \cite{fiebig2005revival}. In spintronics, the internal spin structure protects the magnetic data from external perturbations, which motivates the fast-developing AFM spintronics \cite{jungwirth2016antiferromagnetic}.

Compared to its prominent role in areas above, AFM has not featured prominently in the field of topological materials. However, recent theoretical works \cite{essin2009magnetoelectric, mong2010antiferromagnetic, li2013coupling, gao2014field, tang2016dirac, sivadas2016gate, wang2016generation, Zhang2019a, armitage2019matter, vsmejkal2020crystal, xu2020high, cheong2020seeing, Du2020a, wang2020giant, fei2020giant} have increasingly recognized this exciting prospect. A multitude of fundamentally new topological phenomena that uniquely arise from the interplay between AFM and topology, including the condensed matter realization of Axions (a Dark matter candidate) \cite{li2010dynamical}, the generation of dissipationless spin current in the absence of a concomitant charge current \cite{wang2016generation}, as well as the presence of giant/quantized magneto-electric and magneto-optoelectronic couplings \cite{essin2009magnetoelectric, Zhang2019a, armitage2019matter, wang2020giant, fei2020giant}, have been proposed. Further integrating these new topological physics with spintronics can give rise to the topological AFM spintronics \cite{vsmejkal2018topological}, where the storage, transportation and manipulation of magnetic data can become much faster, more robust and energy-efficient. 

To experimentally fulfill the great potential, crucial open questions, including how topology and Berry curvature are encoded in AFMs and how they can be controlled, remain to be answered. In ferromagnetic topological states, the spontaneous AHE has been widely used to detect the total Berry curvature \cite{nagaosa2010anomalous}. By contrast, in AFMs, the existence of spontaneous AHE is extremely rare. Hence, finding AHE in AFMs, by itself, is of great interest. A primary example is the recent breakthrough identifying the sizable spontaneous AHE in non-collinear antiferromagnets Mn$_3$X (X=Sn or Ge) \cite{chen2014anomalous, nakatsuji2015large, nayak2016large}. The sizable spontaneous AHE of Mn$_3$X therefore directly measures its total Berry curvature. However, as discussed above, the primary characteristic of AFMs is its internal structure. This raises highly intriguing possibilities beyond the spontaneous AHE and the total Berry curvature. We can ask whether topology and Berry phase can acquire distinct spatial structures \cite{li2013coupling, gao2014field} and whether such spatially-locked Berry phase can give rise to new kinds of Hall responses. Here, we investigate these exciting possibilities in even-layered MnBi$_2$Te$_4$ flakes. We report the observation of a fundamentally new type of Hall effect, the layer Hall effect. 

\begin{figure*}[t]
\includegraphics[width=14cm]{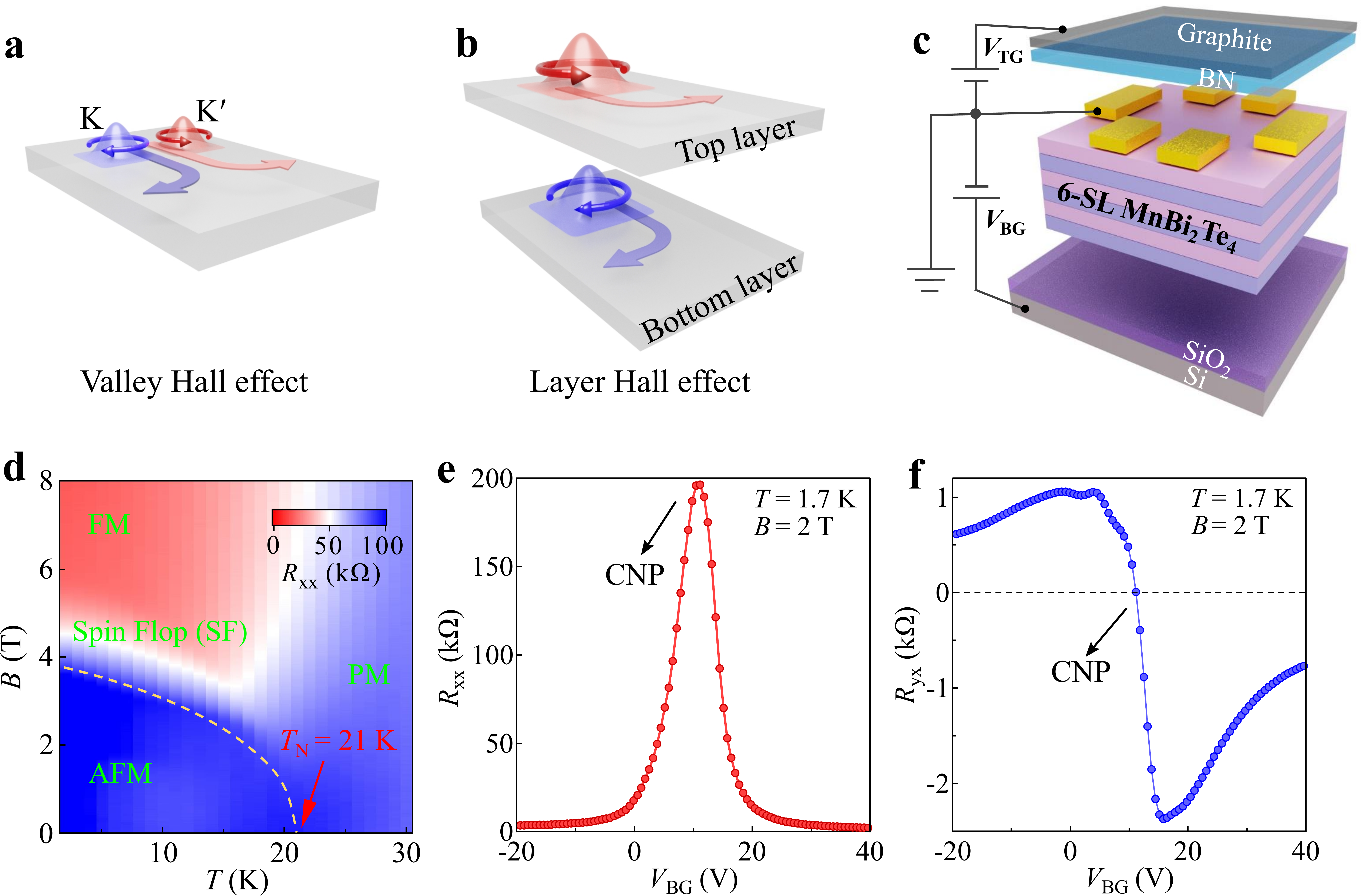}
\caption{{\bf Basic characterizations of the antiferromagnetic 6 septuple layers MnBi$_2$Te$_4$.} \textbf{a,} Illustration of the valley Hall effect. In certain nonmagnetic quantum materials such as gapped graphene and transition metal dichalcogenides (TMDs), Berry curvature is locked to $K$ and $K'$ valleys. Hence electrons of opposite valleys deflect in opposite directions, leading to the valley Hall effect. \textbf{b,} Illustration of the layer Hall effect. In the AFM topological insulator state of even-layered MnBi$_2$Te$_4$, Berry curvature is locked to the top and bottom layers. Hence electrons in the top and bottom layers deflect in opposite directions, leading to the layer Hall effect. A bilayer system is adapted for simplicity. \textbf{c,} Schematic drawing of our dual-gated devices. \textbf{d,} Longitudinal resistance $R_{xx}$ as a function of temperature ($T$) and magnetic field ($B$) with the magnetic states denoted on the data. \textbf{e,f,} $R_{xx}$ and $R_{yx}$ as a function of $V_{\textrm{BG}}$ in the AFM phase at $B=2$ T. $R_{xx}$ as a function of $V_{\textrm{BG}}$ shows a sharp peak. $R_{yx}$ changes sign across charge neutrality.}
\label{Fig1}
\end{figure*}

Beyond charge, electrons can feature additional degrees of freedom such as spin, valley and layer. Encoding Berry curvature with these novel degrees of freedom may lead to new types of Hall effect. A primary example is the valley Hall effect (Fig.~\ref{Fig1}\textbf{a}), which has been realized in gapped graphene and transition metal dichalcogenides \cite{xiao2010berry}. The layer Hall effect studied here is a novel phenomenon where electrons from the top and bottom layers deflect in opposite directions (Fig.~\ref{Fig1}\textbf{b}). 

MnBi$_2$Te$_4$ has recently attracted great interest \cite{Otrokov2019a, Rienks2019, Lee2019a, Yan2019,  Zhang2020b, Zhang2019a, Gordon2019, Chen2019, Hao2019, Swatek2020a, Deng2020, Liu2020a, Ge2020, Liu2020b, Deng2020a, Ovchinnikov2020} since it bridges topology, magnetism and 2D van der Waals (vdW) materials. Its crystal lattice consists of septuple layers (SLs) separated by vdW gaps. Each SL contains seven atomic layers in the sequence of Te-Bi-Te-Mn-Te-Bi-Te. Its magnetic ground state is A-type AFM, where Mn spins within each SL are ferromagnetically aligned along the $z$ axis but Mn spins between adjacent SLs are anti-parallel. Moreover, applying a magnetic field along the $z$ axis can drive the AFM first to a spin-flop phase ($\gtrsim 4$ T) and then to a fully polarized ferromagnetic phase ($\gtrsim 8$ T) \cite{Lee2019a,Yan2019}. Therefore, the magnetic and topological states of 2D MnBi$_2$Te$_4$ flakes can be categorized into two kinds. The first kind is ferromagnetic or ferromagnetic-like, where the 2D system has unequal number of up-spin-layers and down-spin-layers and therefore features an obvious global magnetization. These systems break $\mathcal{PT}$ symmetry. They include odd-layered and even-layered MnBi$_2$Te$_4$ flakes under high magnetic fields as well as odd-layered MnBi$_2$Te$_4$ flakes near zero magnetic field. In these ferromagnetic or ferromagnetic-like systems, pioneering works \cite{Deng2020, Liu2020a, Ge2020, Liu2020b, Deng2020a, Ovchinnikov2020} have reported large/quantized AHE and consistently demonstrated the topological Chern insulator state under high magnetic field. On the other hand, the second kind is the fully-compensated AFM, where the 2D system has an equal number of up-spin-layers and down-spin-layers. These systems preserve $\mathcal{PT}$ symmetry. They include even-layered AFM MnBi$_2$Te$_4$, which are the focus here. In particular, the even-layered MnBi$_2$Te$_4$ is expected to realize the Axion insulator phase, an exotic yet highly rare AFM topological state. Previous works \cite{mogi2017tailoring,xiao2018realization,Liu2020a, Deng2020} have prepared the condition for this phase and observed a highly insulating behavior with zero Hall conductivity. However, the topological and Berry phase properties of the Axion insulator have not been directly probed. We will first focus on presenting the data. Then we will explain how these data demonstrate the layer Hall effect.

We have fabricated high-quality, dual-gated MnBi$_2$Te$_4$ devices (Fig.~\ref{Fig1}\textbf{c}). The dual gating, which has not been explored in MnBi$_2$Te$_4$ especially in even-layered devices \cite{Deng2020, Liu2020a, Ge2020, Liu2020b, Deng2020a, Ovchinnikov2020,Zhang2020b}, is crucial for our experimental findings here, as it allows independent control of the charge density $n$ and the out-of-plane electric field $E$. Particularly, we can set the charge density $n$ at a desired value and study how the system behaves as we only vary the electric field. By contrast, with a single gate, $n$ and $E$ are intrinsically coupled, making the studies of electric field dependence impractical. Moreover, to address the sensitive chemical nature of 2D MnBi$_2$Te$_4$ flakes, we have adapted a high-resolution stencil mask technique (see Methods), enabling us to complete all fabrication processes in an argon environment without exposure to air, chemicals, or heat, therefore preserving the intrinsic nature of the samples. In the main text, we focus on the behavior of a 6SL MnBi$_2$Te$_4$ device (see Extended Data Fig.~\ref{Extended_Data_Figure_QAHE}\textbf{a} for the device image). The overall magnetic behaviors show the expected phase diagram (Fig.~\ref{Fig1}\textbf{d}), including the high-temperature paramagnetic state, the low-temperature AFM ground state, and the magnetic-field-induced ferromagnetic state; (2) The magnetic-field-induced ferromagnetic state is found to be a topological Chern insulator as demonstrated by the vanishing longitudinal resistance $R_{xx}$ and fully ($100\%$) quantized Hall resistance $R_{yx}$ (Extended Data Figs.~\ref{Extended_Data_Figure_QAHE}\textbf{b-d}). These behaviors are all consistent with previous works \cite{Deng2020, Liu2020a, Ge2020, Liu2020b, Deng2020a, Ovchinnikov2020}. 

\begin{figure*}[t]
\includegraphics[width=12.2cm]{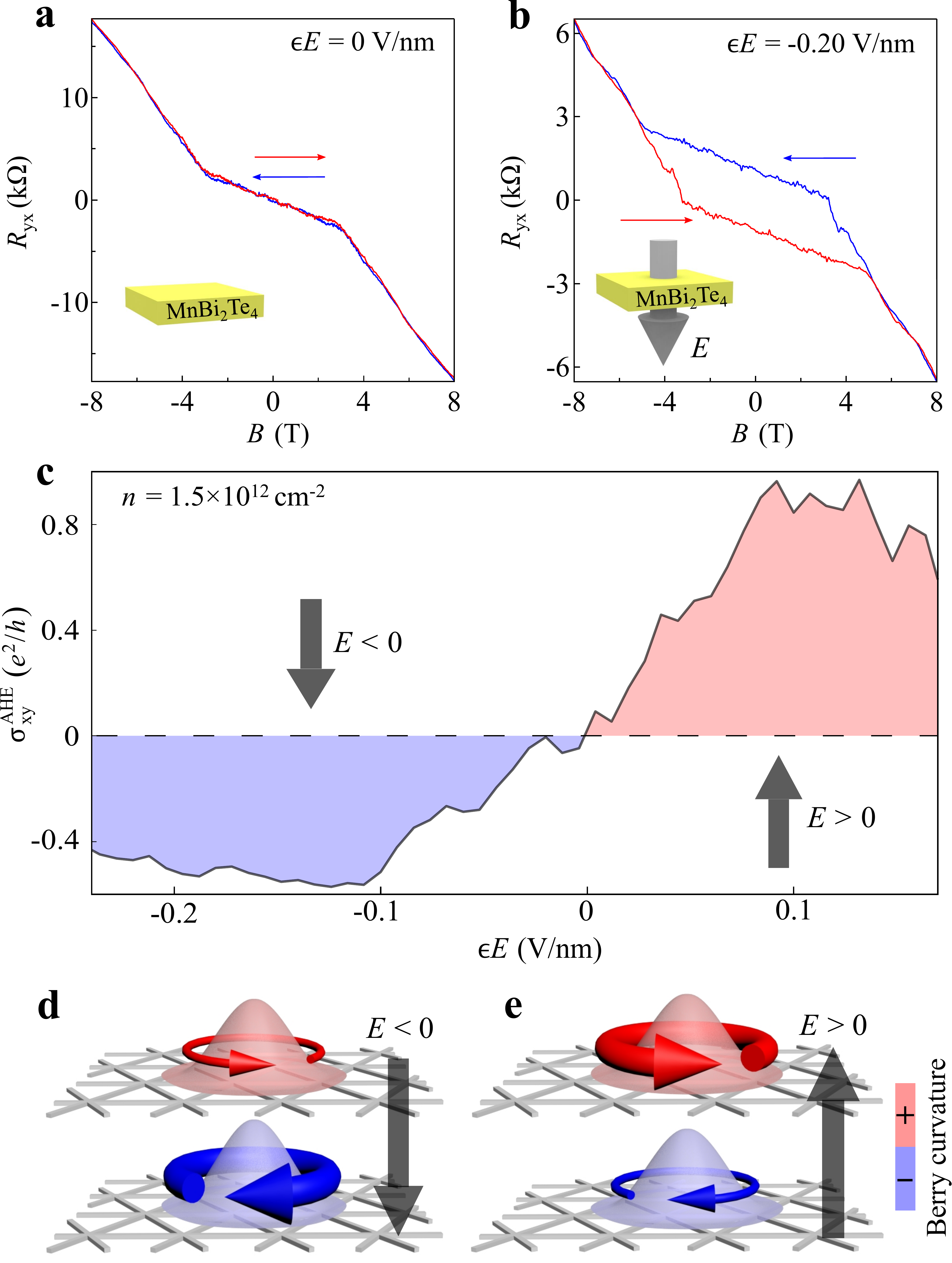}
\caption{{\bf Observation of the layer Hall effect.} \textbf{a,} $R_{yx}$ as a function of $B$ field at zero electric field. Red and blue curves denote the forward and backward scans, respectively. The $E=0$ condition is determined by the maximum of the resistance $R_{xx}$. \textbf{b,} Same as panel (\textbf{a}) but at a finite electric field $\epsilon E=-0.20$ V$/$nm, demonstrating the $E$-field induced AHE. \textbf{c,} The AHE conductivity (at $B=0$) $\sigma_{xy}^{\textrm{AHE}}$ as a function of $E$ field. The charge density $n$ is set in the electron-doped regime ($n=+1.5\times10^{12}$ cm$^{-2}$). \textbf{d,e,} Illustration of the layer-locked Berry curvature under opposite $E$ fields. Depending on the $E$ field direction, the Berry curvature contribution from a particular layer dominates. This explains the AHE's sign reversal with respect to $E$ field. The color denotes the sign of the Berry curvature. The size of the rotating arrows denote the magnitude of the total Berry curvature from a particular layer. A bilayer system is adapted for simplicity.}
\label{Fig2}
\end{figure*}

We now focus on the AFM ground state. The longitudinal resistance $R_{xx}$ (Fig.~\ref{Fig1}\textbf{e}) as a function of the backgate voltage $V_{\textrm{BG}}$ peaks sharply, signaling insulating behavior at charge neutrality. The Hall resistance $R_{yx}$ (Fig.~\ref{Fig1}\textbf{f}) features a clear sign-reversal also at charge neutrality. We note that, in order to eliminate the mixing of $R_{xx}$ and $R_{yx}$ signals, the $R_{yx}$ data in the main text are anti-symmetrized following conventions widely-established in the community \cite{nagaosa2010anomalous}. These $R_{xx}$ and $R_{yx}$ data allow us to obtain the carrier mobility using two independent methods, the field effect model and Hall effect model. Both methods consistently yield a carrier mobility $\sim 1.1\times10^3$ cm$^2/$V$\cdot$s, which is among the highest reported in MnBi$_2$Te$_4$ \cite{Deng2020, Liu2020a, Ge2020, Liu2020b, Deng2020a, Ovchinnikov2020}, highlighting the importance of preserving the intrinsic nature during fabrication. We further study the magnetic-field dependence of $R_{yx}$. As shown in Fig.~\ref{Fig2}\textbf{a}, in the AFM phase, the $R_{yx}$ depends linearly on $B$ field and crosses zero at $B=0$; The forward and backward scans entirely overlap. These observations confirm that the AFM phase does not support any spontaneous AHE, revealing a vanishing total Berry curvature in even-layered MnBi$_2$Te$_4$. As such, the transport measurements above do not show distinct evidence for nontrivial topology/Berry curvature in the AFM phase of 6SL MnBi$_2$Te$_4$. In fact, the observed diverging resistance and vanishing AHE are common behaviors of any nonmagnetic semiconductors. A new experimental tuning parameter is needed to uniquely uncover the rich Berry curvature properties in this AFM topological system.

We now explore the electric field dependence of the AFM phase. We turn on a finite electric field and repeat the measurement of Fig.~\ref{Fig2}\textbf{a}, holding the charge density $n$ and other conditions unchanged. Strikingly, as shown in Fig.~\ref{Fig2}\textbf{b}, we observe that the forward and backward scans become clearly separated in the AFM phase, suggesting the emergence of a large AHE upon the application of electric field. We now study how this AHE depends on $E$ field. Figure~\ref{Fig2}\textbf{c} shows the AHE conductance at $B=0$ as a function of $E$ field while keeping the charge density $n$ at a fixed value in the electron-doped regime ($n=+1.5\times10^{12}$ cm$^{-2}$). Remarkably, we found that the AHE flips sign as the direction of the $E$ field switches. We also change the charge density to the hole-doped regime ($n=-1.4\times10^{12}$ cm$^{-2}$) and repeat the measurement (Extended Data Fig.~\ref{Extended_Data_Figure_Sigma_E_Calculation}\textbf{a}), again, the sign-reversal with respect to $E$ field is observed. We highlight that such an electric-field-reversible AHE has not been achieved previously. It also represents the most fundamental piece of evidence for the layer Hall effect.

\begin{figure*}[t]
\includegraphics[width=15cm]{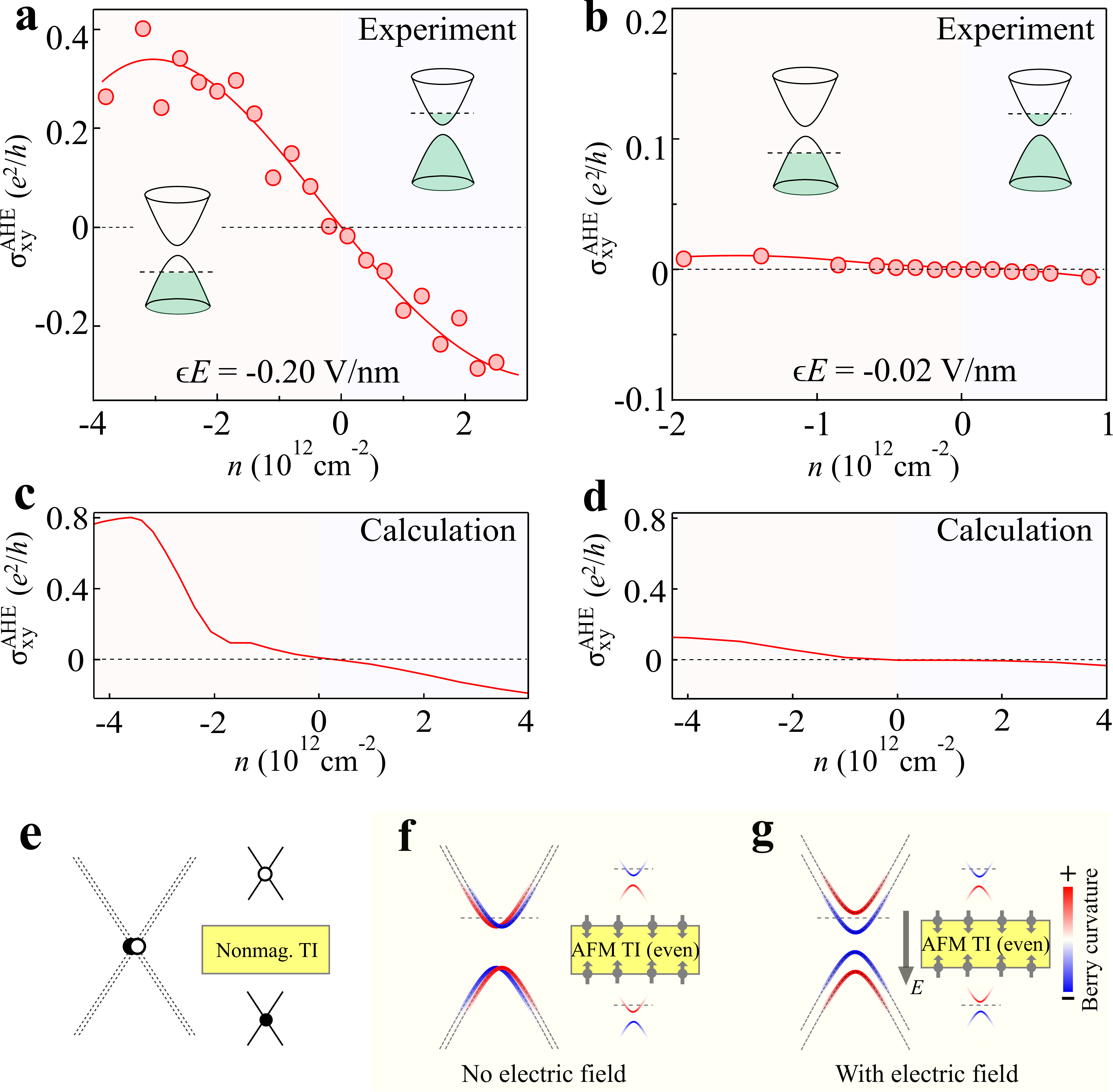}
\caption{{\bf Charge density dependence of the layer Hall effect and the layer-locked Berry curvature.} \textbf{a,b,} The layer Hall effect as a function of charge density $n$ with the $E$ field fixed at $\epsilon E=-0.20$ V$/$nm (panel \textbf{a}) and $\epsilon E=-0.02$ V$/$nm (panel \textbf{b}). The layer Hall effect is observed to show opposite sign in the electron-doped and hole-doped regimes. \textbf{c,d,} First-principles calculated AHE conductivity $\sigma_{xy}^{\textrm{AHE}}$ as a function of charge density. The theoretically applied electric field ($E^{\textrm{THY}}$) here can be related to the displacement field by $D^{\textrm{THY}}=\epsilon^{\textrm{MnBi}_2\textrm{Te}_4} E^{\textrm{THY}}$. \textbf{e-g,} A microscopic picture for the layer Hall effect in 6SL AFM MnBi$_2$Te$_4$. \textbf{e,} A nonmagnetic topological insulator features massless surface Dirac fermions on its top and bottom layers. \textbf{f,} The inclusion of the A-type AFM order gaps the Dirac fermions. The resulting Berry curvature of the top Dirac fermion exactly cancels that of the bottom Dirac fermion. \textbf{g,} Applying an $E$ field can break the degeneracy between the top and bottom Dirac fermions, leading to a large layer-polarized AHE, as long as the Fermi level is away from the band gap. The AHE in the electron-doped regime and in the hole-doped regime has opposite sign.}
\label{Fig3}
\end{figure*}

We then study how the AHE depends on the charge density $n$. Figure~\ref{Fig3}\textbf{a} shows the AHE conductance as a function of $n$ while keeping $E$ field fixed. Interestingly, the AHE is found to show opposite signs depending on whether the system is electron-doped or hole-doped. On the other hand, at charge neutrality, the AHE vanishes. We repeat the measurement with smaller $E$. As shown in Fig.~\ref{Fig3}\textbf{b}, we observe a similar dependence on $n$, although the overall magnitude of the AHE reduces significantly. In Extended Data Figs.~\ref{Extended_Data_Figure_5SL}\textbf{a,b}, we present results on odd-layered MnBi$_2$Te$_4$ devices. As described above, odd-layered MnBi$_2$Te$_4$ is ferromagnetic-like because it has different number of up-spin layers and down-spin layers. As such, it naturally supports AHE without $E$ field. As shown in Extended Data Figs.~\ref{Extended_Data_Figure_5SL}\textbf{a,b}, the AHE in odd-layered MnBi$_2$Te$_4$ remains the same sign as one changes the system from hole-doped to electron-doped, consistent with previous results on odd-layered MnBi$_2$Te$_4$ \cite{Deng2020} and on ferromagnetic Cr- and V-doped topological insulator films \cite{chang2013experimental, kou2014scale, fan2014magnetization, chang2015high, tokura2019magnetic}. The dependence on charge density of the $E$-field-induced AHE in even-layered AFM MnBi$_2$Te$_4$, combined with the distinctly different behavior in odd-layered ferromagnetic-like systems, provides important insights into its microscopic origin. Moreover, the AHE's dependence on charge density also demonstrates that the AHE is not a direct measurement of the magnetization \cite{nagaosa2010anomalous, xiao2010berry, matsukura2015control} (see VI.6 for detailed discussion).

We further study the temperature dependence of the AHE. As shown in Extended Data Fig.~\ref{Extended_Data_Figure_Temperature_Dependence}\textbf{b}, at $T=2$ K, the AHE is pronounced. As we increase temperature to 14 K, the AHE persists but weakens. As we further increase temperature to be above the N\'eel temperature ($T_\textrm{N}\sim21$ K), the AHE is found to vanish entirely. Such temperature dependence demonstrates that the observed AHE is a unique response of the magnetically ordered state.

We enumerate here the key essential data, including the absence of the AHE without $E$ field, the emergence of the AHE upon the application of $E$ field, the AHE's sign-reversal with respect to $E$ field, its unique charge density dependence, its temperature dependence, as well as its sharp contrast with respect to the AHE in odd-layered systems. These data are crucial for excluding various extrinsic effects. For instance, defects and inhomogeneities formed during crystal growth may cause one particular Mn-layer to have a much stronger (or weaker) magnetization than the other Mn layers, thereby breaking the full compensation between up-spins and down-spins. This would lead to a finite magnetization and in turn a spontaneous AHE. In that case, the system should be viewed effectively as a (weak) ferromagnet rather than a fully-compensated AFM. However, the sign of the AHE in a ferromagnet is uniquely tied to the direction of the global magnetization, which cannot be significantly altered or switched by an out-of-plane electric field. Therefore, the AHE of a ferromagnet should not respond dramatically to the electric field, let alone reversing its sign when the electric field is flipped.

We now consider the possibility that the observed AHE is the intrinsic response of the topological AFM phase in 6SL MnBi$_2$Te$_4$. First, we investigate this possibility from a global symmetry point of view. It has been established that the AHE conductivity arises from the total Berry curvature integrated over all occupied state \cite{nagaosa2010anomalous}, i.e., $\sigma^{\textrm{AHE}}=\int_{\textrm{occupied states}} \Omega$, where $\Omega$ is the Berry curvature. Importantly, both $\mathcal{T}$ (time-reversal symmetry) and $\mathcal{PT}$ (the combination of time-reversal and space-inversion symmetries) would force $\sigma^{\textrm{AHE}}$ to vanish. Therefore, for a material to support nonzero AHE, both $\mathcal{T}$ and $\mathcal{PT}$ need to be broken. While all magnetic states break $\mathcal{T}$, what is intriguing about the current AFM is that its internal spin structure preserves $\mathcal{PT}$. Such a $\mathcal{T}$-breaking but $\mathcal{PT}$-symmetric AFM provides the distinct symmetry conditions that uniquely enable an $E$-field-induced AHE: (1) In the absence of $E$ field, the $\mathcal{PT}$ symmetry forces the momentum-integrated Berry curvature and therefore the AHE to vanish; (2) The application of $E$ field breaks $\mathcal{PT}$, which, combined with the fact that $\mathcal{T}$ is already broken, leads to the emergence of a finite AHE; (3) Since the $E$ field provides the critical $\mathcal{PT}$-breaking that results in the nonzero AHE, flipping the $E$ field reverses the way in which $\mathcal{PT}$ is broken, which, in turn, reverses the sign of the AHE. This symmetry analysis is consistent with our experimental observations, providing key evidence for the intrinsic nature.

In order to gain insights beyond the global symmetry analysis, we now examine the microscopic mechanism. As a starting point, a nonmagnetic topological insulator features massless Dirac fermions on its top and bottom surfaces (Fig.~\ref{Fig3}\textbf{e}). Massless Dirac fermions have no Berry curvature ($\Omega=0$) and thus no AHE. The inclusion of a magnetic order gaps these Dirac fermions, leading to giant Berry curvatures in each individual Dirac band (Fig.~\ref{Fig3}\textbf{f}). Due to combination of the A-type AFM spin structure and the even number of layers, the top and bottom Dirac fermions experience opposite magnetizations and hence open up gaps in the opposite fashion (Fig.~\ref{Fig3}\textbf{f}). As such, the Berry curvature contributed by the top Dirac fermion exactly cancels out that contributed by the bottom Dirac fermion, giving rise to a layer-locked Berry curvature. Interestingly, $E$ field can strongly break the degeneracy between top and bottom surfaces. $E>0$ makes the contribution of the bottom layer more dominant, whereas $E<0$ makes the contribution of the top layer more dominant. In other words, the direction of the $E$ field allows us to selectively probe the contribution from one layer over the opposite.

We now further study the dependence on the charge density $n$. We use the charge neutrality with $\sigma^{\textrm{AHE}}=0$ as a reference point. Importantly, we note that the Berry curvature ($\Omega$) of the lowest conduction band and that of the highest valence band are of the same sign in our cases (Fig.~\ref{Fig3}\textbf{g}). Therefore, going from charge neutrality to the electron-doped regime, we add a certain $\Omega$; Going from charge neutrality to the hole-doped regime, we remove the same $\Omega$. Consequently, under the same $E$ field, the layer-polarized AHE in the electron-doped regime and in the hole-doped regime has opposite sign, which is consistent with our experimental observations. We highlight that the sign-reversal with $n$ is a unique signature of the layer Hall effect in $\mathcal{PT}$-symmetric AFM topological insulator (i.e., our even-layered MnBi$_2$Te$_4$ samples). By contrast, in ferromagnetic or ferromagnetic-like systems where there is already a nonzero magnetization $M$, the AHE has the same sign in both electron-doped and hole-doped regimes (see Extended Data Fig.~\ref{Extended_Data_Figure_5SL}).

We now directly compute the band structure, Berry curvature and $\sigma^{\textrm{AHE}}$ of 6SL MnBi$_2$Te$_4$ using both first-principles calculations and tight-biding model calculations. Figures~\ref{Fig3}\textbf{c,d} and Extended Data Figs.~\ref{Extended_Data_Figure_Sigma_E_Calculation}\textbf{c,d} show the first-principles calculated results, which display good agreement with our experimental data on key aspects including the overall magnitude of the $\sigma^{\textrm{AHE}}$ and $\sigma^{\textrm{AHE}}$'s sign-reversal with respect to $E$ and $n$. We note that the existence of a large layer-polarized AHE arising from a tiny Fermi energy ($\lesssim100$ meV for our experiments) suggests giant Berry curvature `per electron', favoring the intrinsic Berry curvature origin over the extrinsic defect scattering origin for the AHE. This is consistent with the magnetic Dirac fermions, where large Berry curvatures are generated near the edges of the band gap. Additionally, we note that theoretical calculations also capture interesting details of our experimental data, such as the weak non-monotonic behavior at relatively large $E$. Overall, our measurements with dual-gated devices therefore uncover unique gate dependencies, consisting of sign-reversals as a function of both $E$ and $n$. These distinct experimental features provide a large parameter space, which allows us to compare them with theoretical analysis and calculations. The $E$-field reversible AHE, taken together with the layer selectivity of the $E$ field, provides convincing evidence for the layer Hall effect and the layer-locked Berry curvature in MnBi$_2$Te$_4$. Notably, the layer Hall effect is already present at $E=0$, although it only manifests itself upon the application of $E$-field. 

\begin{figure*}[t]
\includegraphics[width=14cm]{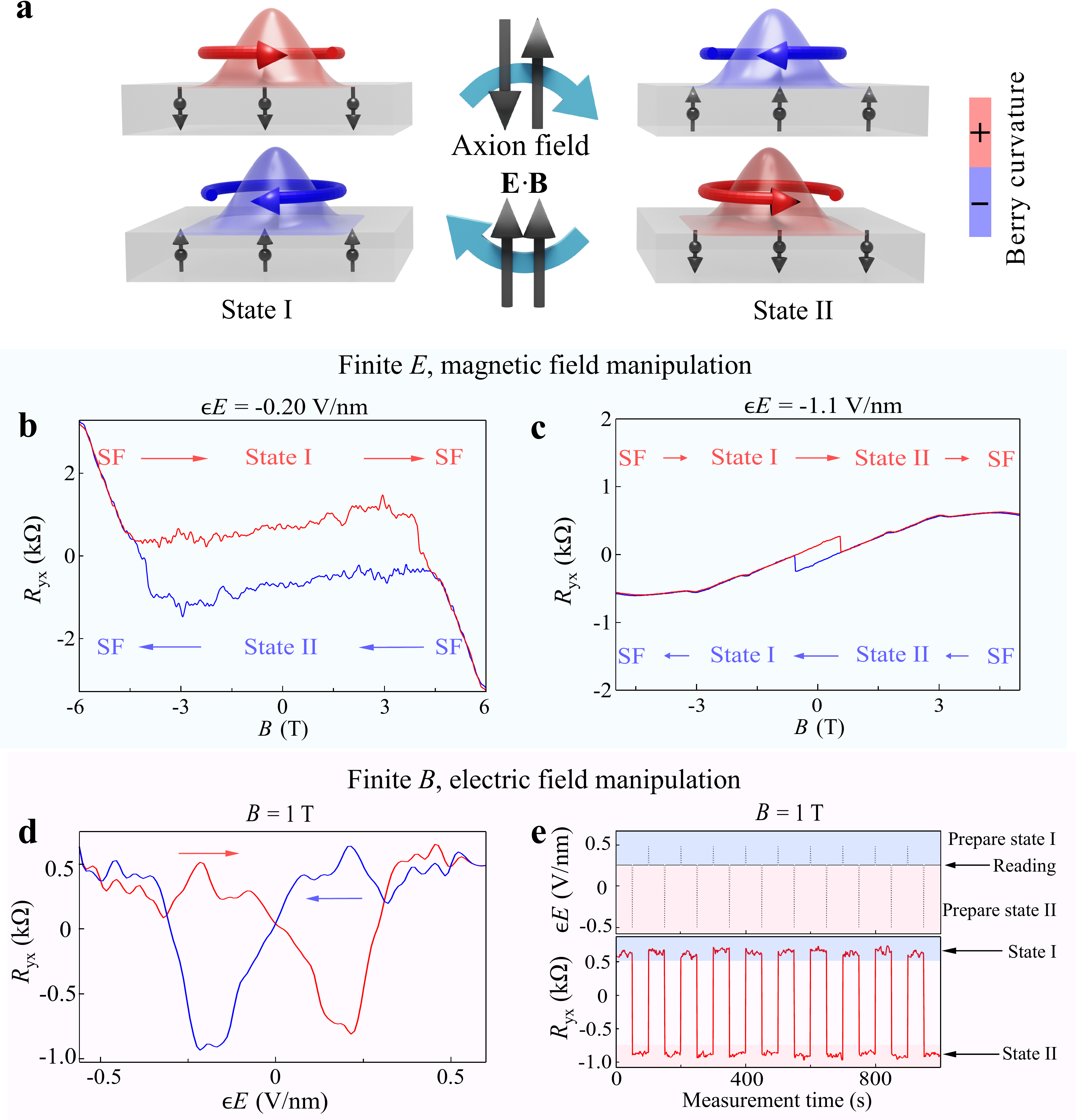}
\caption{{\bf Axion field $\mathbf{E}\cdot\mathbf{B}$ manipulation of the layer-locked Berry curvature and electrical readout by the layer Hall effect.} \textbf{a,} The Axion field $\mathbf{E}\cdot\mathbf{B}$ \cite{fiebig2005revival, li2010dynamical, wang2019dynamic} can switch between the two AFM states (States I and II), which, in turn, reverse the layer-locked Berry curvature. A bilayer system is adapted for simplicity. \textbf{b,} $R_{yx}-B$ with $\epsilon E=-0.20$ V$/$nm. For the forward scan (red curve), at $B\simeq-4$ T, the system changes from the spin-flop phase to the AFM phase with a positive $\mathbf{E}\cdot\mathbf{B}$ ($E<0$ and $B<0$), which favors State I. As we keep moving forward to the $B>0$ region, $\mathbf{E}\cdot\mathbf{B}$ turns negative, which favors State II. However, in this dataset, $\mathbf{E}\cdot\mathbf{B}$ fails to reach the coercive field $(\mathbf{E}\cdot\mathbf{B})_{\textrm{c}}$. Hence state I persists all the way until the system is changed to spin-flop at $B\simeq+4$ T. The situation is similar for the backward scan (blue curve), except that at $B\simeq+4$ T, the system enters the AFM phase with a negative $\mathbf{E}\cdot\mathbf{B}$ field, which favors State II. \textbf{c,} Same as panel (\textbf{b}) but with $\epsilon E=-1.1$ V$/$nm. The much stronger $E$ field leads to a much stronger $\mathbf{E}\cdot\mathbf{B}$ field so that the coercive field $(\mathbf{E}\cdot\mathbf{B})_{\textrm{c}}$ is reached. This induces additional switch between State I and State II at small $B$. \textbf{d,} One can also fix the $B$ field and scan the $E$ field back and forth. Here, $B$ field is fixed at $B=+1$ T. A clear hysteresis as a function of $E$ field is observed. The hysteresis shows a butterfly-like shape, because $E$ field is also needed to generate the AHE, without which there is no contrast between the opposite AFM states. \textbf{e,} By sweeping the $E$ field back and forth, we observe repeatable switching of the AFM states. The $E$ field sweeping is achieved by slow ramping of gate voltage. The measurement time (horizontal axis) only counts the time at the targeted gate voltage. The sweeping time is not shown.}
\label{Fig4}
\end{figure*}

We now explore how one can manipulate the layer-locked Berry curvature. Because the layer-locked Berry curvature is induced by the AFM order, it can be switched by driving the system between the two opposite AFM states (Fig.~\ref{Fig4}\textbf{a}). In ferromagnets, it is well known that one can switch between the two states (magnetization up or down) by $B$ field. Similarly, in ferroelectrics, $E$ field can switch between the electric polarization up and down states. Very interestingly, in our $\mathcal{PT}$-symmetric AFM Axion insulator system, neither $B$ nor $E$ alone can accomplish the switching. Rather,  the Maxwell's equations in MnBi$_2$Te$_4$ are expected to be strongly modified by adding an Axion field term $\Delta L=\theta \frac{e^2}{2\pi hc} \mathbf{E}\cdot\mathbf{B}$ \cite{armitage2019matter, li2010dynamical}, leading to unconventional electromagnetic responses. In particular, the AFM order in the Axion insulator state is expected to couple strongly to the Axion $\mathbf{E}\cdot\mathbf{B}$ field \cite{fiebig2005revival, li2010dynamical, wang2019dynamic}. Although $\Delta L$ can be nonzero in magneto-electric and multiferroic materials \cite{fiebig2005revival, wang2019dynamic} (which are by themselves highly rare and of great interest), there are ample additional interest and merits to explore it in a topological Axion insulator \cite{armitage2019matter} (see discussions below). 

To verify the Axion field $\mathbf{E}\cdot\mathbf{B}$ manipulation, we now examine the hysteresis observed in our $R_{yx}$ $vs$ $B$ data (e.g. Fig.~\ref{Fig4}\textbf{b}) focusing on the two AFM states. We first study the forward scan (red curve) in Fig.~\ref{Fig4}\textbf{b}. Starting from $-8$ T, we increase $B$ field, which controls the magnetism of the system. At $B\simeq-4$ T, the system changes from the spin-flop phase to the AFM phase. On a separate front, a constant, negative $E$ field is always applied in Fig.~\ref{Fig4}\textbf{b}. Therefore, at $B\simeq-4$ T, the system enters the AFM phase with a positive $\mathbf{E}\cdot\mathbf{B}$ field ($E<0$ and $B<0$). This $\mathbf{E}\cdot\mathbf{B}$ field can favor one AFM state over the other. Similarly, for the backward scan (blue curve), the system changes from the spin-flop phase to the AFM phase at $B\simeq+4$ T. In that case, a negative $\mathbf{E}\cdot\mathbf{B}$ is applied near the transition, therefore favoring the opposite AFM state. We note that this control is achieved by going through the meta-magnetic transition from the spin-flop phase to AFM phase with a finite $\mathbf{E}\cdot\mathbf{B}$. As a result, the system changes from spin-flop to a particular AFM state, rather than to both AFM states with equal probability.

To confirm this interpretation, we continue to discuss Fig.~\ref{Fig4}\textbf{b}. For the forward scan in Fig.~\ref{Fig4}\textbf{b}, when the system enters the AFM phase at $B\simeq-4$ T, it experiences a positive $\mathbf{E}\cdot\mathbf{B}$ field, which favors AFM state I. As we continue to scan forward, from $-4$ T to $0$ T, the $\mathbf{E}\cdot\mathbf{B}$ field remains positive. However, as we move to the $B>0$ region, $\mathbf{E}\cdot\mathbf{B}$ turns negative, which favors state II. This negative $\mathbf{E}\cdot\mathbf{B}$ will continue to grow stronger as $B$ increases from $0$ T to $+4$ T. At this point, whether state I can be switched to state II depends on whether $\mathbf{E}\cdot\mathbf{B}$ can reach the coercive field $(\mathbf{E}\cdot\mathbf{B})_{\textrm{c}}$. Interestingly, in Fig.~\ref{Fig4}\textbf{b}, the forward scan shows no sign of switching from state I to II, suggesting that $\mathbf{E}\cdot\mathbf{B}$ fails to reach the coercive field $(\mathbf{E}\cdot\mathbf{B})_{\textrm{c}}$ all the way until the system is changed from AFM to spin-flop at $B\simeq+4$ T. This allows us to make a prediction: If we can repeat the measurement but apply a stronger $E$ field, then we can expect to reach the coercive field $(\mathbf{E}\cdot\mathbf{B})_{\textrm{c}}$ at a lower $B$ field. Indeed, this is confirmed by the measurement shown in Fig.~\ref{Fig4}\textbf{c}, where the switching between the AFM state I and AFM state II are observed in both forward and backward scans, beyond which the two scans overlap. This confirms our prediction and provides a powerful validation of our understanding. Therefore, we have demonstrated that the Axion field $\mathbf{E}\cdot\mathbf{B}$ can switch between the AFM states. This, in turn, manipulates the layer-locked Berry curvature, which is monitored by the layer Hall effect. 

Such an unconventional pseudo-scalar field paves the way for versatile electrical manipulation. In Fig.~\ref{Fig4}\textbf{d}, we fix the $B$ field at $B=+1$ T and scan the $E$ field back and forth. Remarkably, we observe clear hysteresis as a function of $E$ field, which shows a rather unconventional butterfly-like shape. This is because the $E$ field has two important roles: First, $E$ and $B$ together provide the $\mathbf{E}\cdot\mathbf{B}$ field that manipulates the AFM states; Second, $E$ is needed to generate the layer-polarized AHE. As such, we can consistently understand the distinct shape of the hysteresis. In Figs.~\ref{Fig4}\textbf{b,c}, we scan $B$. Because $E$ field is fixed, a finite AHE is always generated. This leads to a square-like hysteresis loop where the forward and backward scans are fully separated near $B=0$. In sharp contrast, in Fig.~\ref{Fig4}\textbf{d}, we scan $E$ with a fixed $B$. At $E=0$, even though a distinct AFM state is favored, the layer-polarized AHE vanishes because of the absence of $E$ field. This leads to a butterfly-like hysteresis where the forward and backward scans coincide at $E=0$. Therefore, the observation of clear hysteresis with respect to $E$ and its butterfly shape provide evidence for electrical switching. By sweeping $E$ back and forth, repeatable switching is observed Fig.~\ref{Fig4}\textbf{e}. 

While our main goal here is to manipulate the layer-locked Berry curvature by the Axion field $\mathbf{E}\cdot\mathbf{B}$, we note that the electrical detection and control of the AFM order by itself is at the core of modern research in spintronics, magnetoelectrics, and multiferroics \cite{fiebig2005revival, jungwirth2016antiferromagnetic}. The prospect of controlling topological magnets is particularly exciting. In our work, first, the Axion field and the layer Hall effect serve as new manipulation and electrical readout methods for the rich internal spatial structures of novel topological AFMs with fully-compensated magnetism. This was not possible previously. Second, we achieve electric control with minimal energy dissipation while retaining the ability to pass a current, offering a new regime that combines the advantages of insulating \cite{fiebig2005revival, matsukura2015control, huang2017layer, jiang2018electric} and metallic \cite{jungwirth2016antiferromagnetic, matsukura2015control, liu2012spin, tsai2020electrical} spintronics. Third, achieving $\mathbf{E}\cdot\mathbf{B}$ control locally will allow us to engineer topological AFM domain wall and Berry curvature moir\'e lattice (Extended Data Fig.~\ref{Extended_Data_Figure_Future}). These novel topological properties are absent in conventional magneto-electric materials. 

Moving forward, our observations suggest many exciting possibilities. First, the layer Hall effect uncovers a unique topological response of fully-compensated AFM Axion insulators, which can be used to distinguish from other materials. For instance, noncentrosymmetric topological materials feature the valley Hall effect \cite{mak2014valley} or the nonlinear Hall effect \cite{sodemann2015quantum}, ferromagnetic topological materials feature the AHE \cite{nagaosa2010anomalous}, whereas the fully-compensated AFM Axion insulators feature the layer Hall effect. Second, the layer Hall effect allows us to understand how Berry phase manifests itself in a fully-compensated AFM with unique Axion topology: Unprecedentedly, Berry curvature is found to acquire a spatial texture. This realization represents the first step towards spatial engineering of Berry phase, such as through layer-specific moi\'e potential (Extended Data Fig.~\ref{Extended_Data_Figure_Future}\textbf{b}). The spatially-locked Berry curvature can significantly modify the electrical, spin, optical, and optoelectronic properties of topological AFMs, giving rise to novel effects \cite{li2013coupling, sivadas2016gate, wang2016generation, wang2020giant, fei2020giant} such as $E$-field-induced magneto-optical effects, dissipationless spin currents and topological AFM domain wall modes. Third, our demonstration of the Axion field $\mathbf{E}\cdot\mathbf{B}$ manipulation of the AFM states indicates strong magneto-electric coupling in even-layered MnBi$_2$Te$_4$. Although the magneto-electric coupling is also present in conventional magneto-electric and multiferroic materials such as Cr$_2$O$_3$ \cite{fiebig2005revival, Iyama_2013} and CrI$_3$ \cite{huang2017layer, jiang2018electric}, MnBi$_2$Te$_4$ is the first magneto-electric system that supports magneto-electric control and nontrivial topology simultaneously. Moreover, conventional magneto-electrics are wide-gap magnetic insulators that do not support electrical transport at all. By contrast, MnBi$_2$Te$_4$  features gate tunable charge transport with novel $E$-field induced Hall effect. Further, the magneto-electric coupling in those conventional magneto-electrics mainly arise from localized magnetic ions. By contrast, the magneto-electric coupling in MnBi$_2$Te$_4$ is expected to be dominated by the contribution from low-energy topological electrons, leading to the novel quantized Axion coupling \cite{essin2009magnetoelectric, Zhang2019a, armitage2019matter}. Finally, the AFM domain walls in even-layered MnBi$_2$Te$_4$ are predicted to support layer-spacific topological modes \cite{li2013coupling} (Extended Data Fig.~\ref{Extended_Data_Figure_Future}\textbf{a}). By achieving local control of the AFM domains using piezoelectric or magnetic force tips based on the Axion field $\mathbf{E}\cdot\mathbf{B}$ manipulation, one can design networks of topological conduction channels. Versatile controls of the layer-locked Berry curvatures, as begun in our work, can result in the exciting applications in topological AFM spintronics.

\vspace{0.5cm}

\bibliographystyle{naturemag}
\bibliography{Topological_and_2D_11102020}

\begin{thebibliography}{10}
\expandafter\ifx\csname url\endcsname\relax
  \def\url#1{\texttt{#1}}\fi
\expandafter\ifx\csname urlprefix\endcsname\relax\def\urlprefix{URL }\fi
\providecommand{\bibinfo}[2]{#2}
\providecommand{\eprint}[2][]{\url{#2}}

\bibitem{neel1972magnetism}
\bibinfo{author}{N\'eel, L.}
\newblock \bibinfo{title}{Magnetism ans the local molecular field}.
\newblock \emph{\bibinfo{journal}{Nobel Lectures, Physic}}
  (\bibinfo{year}{1970}).

\bibitem{li2013coupling}
\bibinfo{author}{Li, X.}, \bibinfo{author}{Cao, T.}, \bibinfo{author}{Niu, Q.},
  \bibinfo{author}{Shi, J.} \& \bibinfo{author}{Feng, J.}
\newblock \bibinfo{title}{Coupling the valley degree of freedom to
  antiferromagnetic order}.
\newblock \emph{\bibinfo{journal}{PNAS}} \textbf{\bibinfo{volume}{110}},
  \bibinfo{pages}{3738--3742} (\bibinfo{year}{2013}).

\bibitem{gao2014field}
\bibinfo{author}{Gao, Y.}, \bibinfo{author}{Yang, S.~A.} \&
  \bibinfo{author}{Niu, Q.}
\newblock \bibinfo{title}{Field induced positional shift of {B}loch electrons
  and its dynamical implications}.
\newblock \emph{\bibinfo{journal}{Phys. Rev. Lett.}}
  \textbf{\bibinfo{volume}{112}}, \bibinfo{pages}{166601}
  (\bibinfo{year}{2014}).

\bibitem{chen2014anomalous}
\bibinfo{author}{Chen, H.}, \bibinfo{author}{Niu, Q.} \&
  \bibinfo{author}{MacDonald, A.}
\newblock \bibinfo{title}{Anomalous {H}all effect arising from noncollinear
  antiferromagnetism}.
\newblock \emph{\bibinfo{journal}{Phys. Rev. Lett.}}
  \textbf{\bibinfo{volume}{112}}, \bibinfo{pages}{017205}
  (\bibinfo{year}{2014}).

\bibitem{nakatsuji2015large}
\bibinfo{author}{Nakatsuji, S.}, \bibinfo{author}{Kiyohara, N.} \&
  \bibinfo{author}{Higo, T.}
\newblock \bibinfo{title}{Large anomalous {H}all effect in a non-collinear
  antiferromagnet at room temperature}.
\newblock \emph{\bibinfo{journal}{Nature}} \textbf{\bibinfo{volume}{527}},
  \bibinfo{pages}{212--215} (\bibinfo{year}{2015}).

\bibitem{nayak2016large}
\bibinfo{author}{Nayak, A.~K.} \emph{et~al.}
\newblock \bibinfo{title}{Large anomalous {H}all effect driven by a
  nonvanishing {B}erry curvature in the noncolinear antiferromagnet
  {M}n$_3${G}e}.
\newblock \emph{\bibinfo{journal}{Science Advances}}
  \textbf{\bibinfo{volume}{2}}, \bibinfo{pages}{e1501870}
  (\bibinfo{year}{2016}).

\bibitem{mong2010antiferromagnetic}
\bibinfo{author}{Mong, R.~S.}, \bibinfo{author}{Essin, A.~M.} \&
  \bibinfo{author}{Moore, J.~E.}
\newblock \bibinfo{title}{Antiferromagnetic topological insulators}.
\newblock \emph{\bibinfo{journal}{Phys. Rev. B}} \textbf{\bibinfo{volume}{81}},
  \bibinfo{pages}{245209} (\bibinfo{year}{2010}).

\bibitem{vsmejkal2018topological}
\bibinfo{author}{{\v{S}}mejkal, L.}, \bibinfo{author}{Mokrousov, Y.},
  \bibinfo{author}{Yan, B.} \& \bibinfo{author}{MacDonald, A.~H.}
\newblock \bibinfo{title}{Topological antiferromagnetic spintronics}.
\newblock \emph{\bibinfo{journal}{Nature Phys.}} \textbf{\bibinfo{volume}{14}},
  \bibinfo{pages}{242--251} (\bibinfo{year}{2018}).

\bibitem{tokura2019magnetic}
\bibinfo{author}{Tokura, Y.}, \bibinfo{author}{Yasuda, K.} \&
  \bibinfo{author}{Tsukazaki, A.}
\newblock \bibinfo{title}{Magnetic topological insulators}.
\newblock \emph{\bibinfo{journal}{Nature Reviews Physics}}
  \textbf{\bibinfo{volume}{1}}, \bibinfo{pages}{126--143}
  (\bibinfo{year}{2019}).

\bibitem{mong2019magnetic}
\bibinfo{author}{Mong, R.~S.} \& \bibinfo{author}{Moore, J.~E.}
\newblock \bibinfo{title}{Magnetic and topological order united in a crystal}.
\newblock \emph{\bibinfo{journal}{Nature}} \textbf{\bibinfo{volume}{576}},
  \bibinfo{pages}{390--392} (\bibinfo{year}{2019}).

\bibitem{xu2020high}
\bibinfo{author}{Xu, Y.} \emph{et~al.}
\newblock \bibinfo{title}{High-throughput calculations of magnetic topological
  materials}.
\newblock \emph{\bibinfo{journal}{Nature}} \textbf{\bibinfo{volume}{586}},
  \bibinfo{pages}{702--707} (\bibinfo{year}{2020}).

\bibitem{cheong2020seeing}
\bibinfo{author}{Cheong, S.-W.}, \bibinfo{author}{Fiebig, M.},
  \bibinfo{author}{Wu, W.}, \bibinfo{author}{Chapon, L.} \&
  \bibinfo{author}{Kiryukhin, V.}
\newblock \bibinfo{title}{Seeing is believing: visualization of
  antiferromagnetic domains}.
\newblock \emph{\bibinfo{journal}{npj Quantum Materials}}
  \textbf{\bibinfo{volume}{5}}, \bibinfo{pages}{1--10} (\bibinfo{year}{2020}).

\bibitem{lee2006doping}
\bibinfo{author}{Lee, P.~A.}, \bibinfo{author}{Nagaosa, N.} \&
  \bibinfo{author}{Wen, X.-G.}
\newblock \bibinfo{title}{Doping a {M}ott insulator: Physics of
  high-temperature superconductivity}.
\newblock \emph{\bibinfo{journal}{Rev. Mod. Phys.}}
  \textbf{\bibinfo{volume}{78}}, \bibinfo{pages}{17--84}
  (\bibinfo{year}{2006}).

\bibitem{fiebig2005revival}
\bibinfo{author}{Fiebig, M.}
\newblock \bibinfo{title}{Revival of the magnetoelectric effect}.
\newblock \emph{\bibinfo{journal}{Journal of Physics D: Applied Physics}}
  \textbf{\bibinfo{volume}{38}}, \bibinfo{pages}{R123--R152}
  (\bibinfo{year}{2005}).

\bibitem{jungwirth2016antiferromagnetic}
\bibinfo{author}{Jungwirth, T.}, \bibinfo{author}{Marti, X.},
  \bibinfo{author}{Wadley, P.} \& \bibinfo{author}{Wunderlich, J.}
\newblock \bibinfo{title}{Antiferromagnetic spintronics}.
\newblock \emph{\bibinfo{journal}{Nature Nanotech.}}
  \textbf{\bibinfo{volume}{11}}, \bibinfo{pages}{231--241}
  (\bibinfo{year}{2016}).

\bibitem{essin2009magnetoelectric}
\bibinfo{author}{Essin, A.~M.}, \bibinfo{author}{Moore, J.~E.} \&
  \bibinfo{author}{Vanderbilt, D.}
\newblock \bibinfo{title}{Magnetoelectric polarizability and axion
  electrodynamics in crystalline insulators}.
\newblock \emph{\bibinfo{journal}{Phys. Rev. Lett.}}
  \textbf{\bibinfo{volume}{102}}, \bibinfo{pages}{146805}
  (\bibinfo{year}{2009}).

\bibitem{tang2016dirac}
\bibinfo{author}{Tang, P.}, \bibinfo{author}{Zhou, Q.}, \bibinfo{author}{Xu,
  G.} \& \bibinfo{author}{Zhang, S.-C.}
\newblock \bibinfo{title}{Dirac fermions in an antiferromagnetic semimetal}.
\newblock \emph{\bibinfo{journal}{Nature Phys.}} \textbf{\bibinfo{volume}{12}},
  \bibinfo{pages}{1100--1104} (\bibinfo{year}{2016}).

\bibitem{sivadas2016gate}
\bibinfo{author}{Sivadas, N.}, \bibinfo{author}{Okamoto, S.} \&
  \bibinfo{author}{Xiao, D.}
\newblock \bibinfo{title}{Gate-controllable magneto-optic {K}err effect in
  layered collinear antiferromagnets}.
\newblock \emph{\bibinfo{journal}{Phys. Rev. Lett.}}
  \textbf{\bibinfo{volume}{117}}, \bibinfo{pages}{267203}
  (\bibinfo{year}{2016}).

\bibitem{wang2016generation}
\bibinfo{author}{Wang, J.}, \bibinfo{author}{Lian, B.} \&
  \bibinfo{author}{Zhang, S.-C.}
\newblock \bibinfo{title}{Generation of spin currents by magnetic field in
  $\mathcal{T}$-and $\mathcal{P}$-broken materials}.
\newblock \emph{\bibinfo{journal}{Spin}} \textbf{\bibinfo{volume}{09}},
  \bibinfo{pages}{04} (\bibinfo{year}{2019}).

\bibitem{Zhang2019a}
\bibinfo{author}{Zhang, D.} \emph{et~al.}
\newblock \bibinfo{title}{Topological axion states in the magnetic insulator
  {M}n{B}i$_2${T}e$_4$ with the quantized magnetoelectric effect}.
\newblock \emph{\bibinfo{journal}{Phys. Rev. Lett.}}
  \textbf{\bibinfo{volume}{122}}, \bibinfo{pages}{206401}
  (\bibinfo{year}{2019}).

\bibitem{armitage2019matter}
\bibinfo{author}{Armitage, N.~P.} \& \bibinfo{author}{Wu, L.}
\newblock \bibinfo{title}{On the matter of topological insulators as
  magnetoelectrics}.
\newblock \emph{\bibinfo{journal}{SciPost Phys.}} \textbf{\bibinfo{volume}{6}},
  \bibinfo{pages}{046} (\bibinfo{year}{2019}).

\bibitem{vsmejkal2020crystal}
\bibinfo{author}{{\v{S}}mejkal, L.},
  \bibinfo{author}{Gonz{\'a}lez-Hern{\'a}ndez, R.}, \bibinfo{author}{Jungwirth,
  T.} \& \bibinfo{author}{Sinova, J.}
\newblock \bibinfo{title}{Crystal time-reversal symmetry breaking and
  spontaneous {H}all effect in collinear antiferromagnets}.
\newblock \emph{\bibinfo{journal}{Science Advances}}
  \textbf{\bibinfo{volume}{6}}, \bibinfo{pages}{eaaz8809}
  (\bibinfo{year}{2020}).

\bibitem{Du2020a}
\bibinfo{author}{Du, S.} \emph{et~al.}
\newblock \bibinfo{title}{Berry curvature engineering by gating two-dimensional
  antiferromagnets}.
\newblock \emph{\bibinfo{journal}{Phys. Rev. Research}}
  \textbf{\bibinfo{volume}{2}}, \bibinfo{pages}{022025} (\bibinfo{year}{2020}).

\bibitem{wang2020giant}
\bibinfo{author}{Wang, H.} \& \bibinfo{author}{Qian, X.}
\newblock \bibinfo{title}{Giant nonlinear photocurrent in
  $\mathcal{PT}$-symmetric magnetic topological quantum materials}.
\newblock \emph{\bibinfo{journal}{arXiv preprint}}
  \bibinfo{pages}{arXiv:2006.13573} (\bibinfo{year}{2020}).

\bibitem{fei2020giant}
\bibinfo{author}{Fei, R.}, \bibinfo{author}{Song, W.} \& \bibinfo{author}{Yang,
  L.}
\newblock \bibinfo{title}{Giant linearly-polarized photogalvanic effect and
  second harmonic generation in two-dimensional axion insulators}.
\newblock \emph{\bibinfo{journal}{arXiv preprint}}
  \bibinfo{pages}{arXiv:2003.01576} (\bibinfo{year}{2020}).

\bibitem{li2010dynamical}
\bibinfo{author}{Li, R.}, \bibinfo{author}{Wang, J.}, \bibinfo{author}{Qi,
  X.-L.} \& \bibinfo{author}{Zhang, S.-C.}
\newblock \bibinfo{title}{Dynamical axion field in topological magnetic
  insulators}.
\newblock \emph{\bibinfo{journal}{Nature Phys.}} \textbf{\bibinfo{volume}{6}},
  \bibinfo{pages}{284--288} (\bibinfo{year}{2010}).

\bibitem{nagaosa2010anomalous}
\bibinfo{author}{Nagaosa, N.}, \bibinfo{author}{Sinova, J.},
  \bibinfo{author}{Onoda, S.}, \bibinfo{author}{MacDonald, A.} \&
  \bibinfo{author}{Ong, N.~P.}
\newblock \bibinfo{title}{Anomalous {H}all effect}.
\newblock \emph{\bibinfo{journal}{Rev. Mod. Phys.}}
  \textbf{\bibinfo{volume}{82}}, \bibinfo{pages}{1539--1592}
  (\bibinfo{year}{2010}).

\bibitem{xiao2010berry}
\bibinfo{author}{Xiao, D.}, \bibinfo{author}{Chang, M.-C.} \&
  \bibinfo{author}{Niu, Q.}
\newblock \bibinfo{title}{Berry phase effects on electronic properties}.
\newblock \emph{\bibinfo{journal}{Rev. Mod. Phys.}}
  \textbf{\bibinfo{volume}{82}}, \bibinfo{pages}{1959--2007}
  (\bibinfo{year}{2010}).

\bibitem{Otrokov2019a}
\bibinfo{author}{Otrokov, M.~M.} \emph{et~al.}
\newblock \bibinfo{title}{Prediction and observation of an antiferromagnetic
  topological insulator}.
\newblock \emph{\bibinfo{journal}{Nature}} \textbf{\bibinfo{volume}{576}},
  \bibinfo{pages}{416--422} (\bibinfo{year}{2019}).

\bibitem{Rienks2019}
\bibinfo{author}{Rienks, E. D.~L.} \emph{et~al.}
\newblock \bibinfo{title}{Large magnetic gap at the {D}irac point in
  {B}i$_2${T}e$_3$/{M}n{B}i$_2${T}e$_4$ heterostructures}.
\newblock \emph{\bibinfo{journal}{Nature}} \textbf{\bibinfo{volume}{576}},
  \bibinfo{pages}{423--428} (\bibinfo{year}{2019}).

\bibitem{Lee2019a}
\bibinfo{author}{Lee, S.~H.} \emph{et~al.}
\newblock \bibinfo{title}{Spin scattering and noncollinear spin
  structure-induced intrinsic anomalous {H}all effect in antiferromagnetic
  topological insulator
  $\mathrm{MnB}{\mathrm{i}}_{2}\mathrm{T}{\mathrm{e}}_{4}$}.
\newblock \emph{\bibinfo{journal}{Phys. Rev. Research}}
  \textbf{\bibinfo{volume}{1}}, \bibinfo{pages}{012011} (\bibinfo{year}{2019}).

\bibitem{Yan2019}
\bibinfo{author}{Yan, J.-Q.} \emph{et~al.}
\newblock \bibinfo{title}{Crystal growth and magnetic structure of
  {M}n{B}i$_2${T}e$_4$}.
\newblock \emph{\bibinfo{journal}{Phys. Rev. Materials}}
  \textbf{\bibinfo{volume}{3}}, \bibinfo{pages}{064202} (\bibinfo{year}{2019}).

\bibitem{Zhang2020b}
\bibinfo{author}{Zhang, S.} \emph{et~al.}
\newblock \bibinfo{title}{Experimental observation of the gate-controlled
  reversal of the anomalous {H}all effect in the intrinsic magnetic topological
  insulator {M}n{B}i$_2${T}e$_4$ device}.
\newblock \emph{\bibinfo{journal}{Nano Lett.}} \textbf{\bibinfo{volume}{20}},
  \bibinfo{pages}{709--714} (\bibinfo{year}{2020}).

\bibitem{Gordon2019}
\bibinfo{author}{Gordon, K.~N.} \emph{et~al.}
\newblock \bibinfo{title}{Strongly gapped topological surface states on
  protected surfaces of antiferromagnetic {M}n{B}i$_4${T}e$ _7$ and
  {M}n{B}i$_6${T}e$_{10}$}.
\newblock \emph{\bibinfo{journal}{arXiv preprint}}
  \bibinfo{pages}{arXiv:1910.13943} (\bibinfo{year}{2019}).

\bibitem{Chen2019}
\bibinfo{author}{Chen, Y.~J.} \emph{et~al.}
\newblock \bibinfo{title}{Topological electronic structure and its temperature
  evolution in antiferromagnetic topological insulator {M}n{B}i$_2${T}e$_4$}.
\newblock \emph{\bibinfo{journal}{Phys. Rev. X}} \textbf{\bibinfo{volume}{9}},
  \bibinfo{pages}{041040} (\bibinfo{year}{2019}).

\bibitem{Hao2019}
\bibinfo{author}{Hao, Y.-J.} \emph{et~al.}
\newblock \bibinfo{title}{Gapless surface {D}irac cone in antiferromagnetic
  topological insulator {M}n{B}i$_2${T}e$_4$}.
\newblock \emph{\bibinfo{journal}{Phys. Rev. X}} \textbf{\bibinfo{volume}{9}},
  \bibinfo{pages}{041038} (\bibinfo{year}{2019}).

\bibitem{Swatek2020a}
\bibinfo{author}{Swatek, P.} \emph{et~al.}
\newblock \bibinfo{title}{Gapless {D}irac surface states in the
  antiferromagnetic topological insulator {M}n{B}i$_2${T}e$_4$}.
\newblock \emph{\bibinfo{journal}{Phys. Rev. B}}
  \textbf{\bibinfo{volume}{101}}, \bibinfo{pages}{161109}
  (\bibinfo{year}{2020}).

\bibitem{Deng2020}
\bibinfo{author}{Deng, Y.} \emph{et~al.}
\newblock \bibinfo{title}{Quantum anomalous {H}all effect in intrinsic magnetic
  topological insulator {M}n{B}i$_2${T}e$_4$}.
\newblock \emph{\bibinfo{journal}{Science}} \textbf{\bibinfo{volume}{367}},
  \bibinfo{pages}{895--900} (\bibinfo{year}{2020}).

\bibitem{Liu2020a}
\bibinfo{author}{Liu, C.} \emph{et~al.}
\newblock \bibinfo{title}{Robust axion insulator and {C}hern insulator phases
  in a two-dimensional antiferromagnetic topological insulator}.
\newblock \emph{\bibinfo{journal}{Nature Mater.}}
  \textbf{\bibinfo{volume}{19}}, \bibinfo{pages}{522--527}
  (\bibinfo{year}{2020}).

\bibitem{Ge2020}
\bibinfo{author}{Ge, J.} \emph{et~al.}
\newblock \bibinfo{title}{High-{C}hern-number and high-temperature quantum
  {H}all effect without landau levels}.
\newblock \emph{\bibinfo{journal}{Natl. Sci. Rev.}}
  \textbf{\bibinfo{volume}{7}}, \bibinfo{pages}{1280--1287}
  (\bibinfo{year}{2020}).

\bibitem{Liu2020b}
\bibinfo{author}{Liu, C.} \emph{et~al.}
\newblock \bibinfo{title}{Helical {C}hern insulator phase with broken
  time-reversal symmetry in {M}n{B}i$_2${T}e$_4$}.
\newblock \emph{\bibinfo{journal}{arXiv preprint}}
  \bibinfo{pages}{arXiv:2001.08401} (\bibinfo{year}{2020}).

\bibitem{Deng2020a}
\bibinfo{author}{Deng, H.} \emph{et~al.}
\newblock \bibinfo{title}{High-temperature quantum anomalous {H}all regime in a
  {M}n{B}i$_2${T}e$_4/${B}i$_2${T}e$_3$ superlattice}.
\newblock \emph{\bibinfo{journal}{Nature Phys.}}  (\bibinfo{year}{2020}).

\bibitem{Ovchinnikov2020}
\bibinfo{author}{Ovchinnikov, D.} \emph{et~al.}
\newblock \bibinfo{title}{Intertwined topological and magnetic orders in
  atomically thin {C}hern insulator {M}n{B}i$_2${T}e$_4$}.
\newblock \emph{\bibinfo{journal}{arXiv preprint}}
  \bibinfo{pages}{arXiv:2011.00555} (\bibinfo{year}{2020}).

\bibitem{mogi2017tailoring}
\bibinfo{author}{Mogi, M.} \emph{et~al.}
\newblock \bibinfo{title}{Tailoring tricolor structure of magnetic topological
  insulator for robust axion insulator}.
\newblock \emph{\bibinfo{journal}{Science Advances}}
  \textbf{\bibinfo{volume}{3}}, \bibinfo{pages}{eaao1669}
  (\bibinfo{year}{2017}).

\bibitem{xiao2018realization}
\bibinfo{author}{Xiao, D.} \emph{et~al.}
\newblock \bibinfo{title}{Realization of the axion insulator state in quantum
  anomalous {H}all sandwich heterostructures}.
\newblock \emph{\bibinfo{journal}{Phys. Rev. Lett.}}
  \textbf{\bibinfo{volume}{120}}, \bibinfo{pages}{056801}
  (\bibinfo{year}{2018}).

\bibitem{chang2013experimental}
\bibinfo{author}{Chang, C.-Z.} \emph{et~al.}
\newblock \bibinfo{title}{Experimental {O}bservation of the {Q}uantum
  {A}nomalous {H}all {E}ffect in a {M}agnetic {T}opological {I}nsulator}.
\newblock \emph{\bibinfo{journal}{Science}} \textbf{\bibinfo{volume}{340}},
  \bibinfo{pages}{167--170} (\bibinfo{year}{2013}).

\bibitem{kou2014scale}
\bibinfo{author}{Kou, X.} \emph{et~al.}
\newblock \bibinfo{title}{Scale-invariant quantum anomalous {H}all effect in
  magnetic topological insulators beyond the two-dimensional limit}.
\newblock \emph{\bibinfo{journal}{Phys. Rev. Lett.}}
  \textbf{\bibinfo{volume}{113}}, \bibinfo{pages}{137201}
  (\bibinfo{year}{2014}).

\bibitem{fan2014magnetization}
\bibinfo{author}{Fan, Y.} \emph{et~al.}
\newblock \bibinfo{title}{Magnetization switching through giant spin-orbit
  torque in a magnetically doped topological insulator heterostructure}.
\newblock \emph{\bibinfo{journal}{Nature Mater.}}
  \textbf{\bibinfo{volume}{13}}, \bibinfo{pages}{699--704}
  (\bibinfo{year}{2014}).

\bibitem{chang2015high}
\bibinfo{author}{Chang, C.-Z.} \emph{et~al.}
\newblock \bibinfo{title}{High-precision realization of robust quantum
  anomalous {H}all state in a hard ferromagnetic topological insulator}.
\newblock \emph{\bibinfo{journal}{Nature materials}}
  \textbf{\bibinfo{volume}{14}}, \bibinfo{pages}{473--477}
  (\bibinfo{year}{2015}).

\bibitem{matsukura2015control}
\bibinfo{author}{Matsukura, F.}, \bibinfo{author}{Tokura, Y.} \&
  \bibinfo{author}{Ohno, H.}
\newblock \bibinfo{title}{Control of magnetism by electric fields}.
\newblock \emph{\bibinfo{journal}{Nature Nanotech.}}
  \textbf{\bibinfo{volume}{10}}, \bibinfo{pages}{209--220}
  (\bibinfo{year}{2015}).

\bibitem{wang2019dynamic}
\bibinfo{author}{Wang, J.}, \bibinfo{author}{Lei, C.},
  \bibinfo{author}{Macdonald, A.~H.} \& \bibinfo{author}{Binek, C.}
\newblock \bibinfo{title}{Dynamic axion field in the magnetoelectric
  antiferromagnet chromia}.
\newblock \emph{\bibinfo{journal}{arXiv preprint}}
  \bibinfo{pages}{arXiv:1901.08536} (\bibinfo{year}{2019}).

\bibitem{huang2017layer}
\bibinfo{author}{Huang, B.} \emph{et~al.}
\newblock \bibinfo{title}{Layer-dependent ferromagnetism in a van der {W}aals
  crystal down to the monolayer limit}.
\newblock \emph{\bibinfo{journal}{Nature}} \textbf{\bibinfo{volume}{546}},
  \bibinfo{pages}{270--273} (\bibinfo{year}{2017}).

\bibitem{jiang2018electric}
\bibinfo{author}{Jiang, S.}, \bibinfo{author}{Shan, J.} \&
  \bibinfo{author}{Mak, K.~F.}
\newblock \bibinfo{title}{Electric-field switching of two-dimensional van der
  {W}aals magnets}.
\newblock \emph{\bibinfo{journal}{Nature Mater.}}
  \textbf{\bibinfo{volume}{17}}, \bibinfo{pages}{406--410}
  (\bibinfo{year}{2018}).

\bibitem{liu2012spin}
\bibinfo{author}{Liu, L.} \emph{et~al.}
\newblock \bibinfo{title}{Spin-torque switching with the giant spin {H}all
  effect of tantalum}.
\newblock \emph{\bibinfo{journal}{Science}} \textbf{\bibinfo{volume}{336}},
  \bibinfo{pages}{555--558} (\bibinfo{year}{2012}).

\bibitem{tsai2020electrical}
\bibinfo{author}{Tsai, H.} \emph{et~al.}
\newblock \bibinfo{title}{Electrical manipulation of a topological
  antiferromagnetic state}.
\newblock \emph{\bibinfo{journal}{Nature}} \textbf{\bibinfo{volume}{580}},
  \bibinfo{pages}{608--613} (\bibinfo{year}{2020}).

\bibitem{mak2014valley}
\bibinfo{author}{Mak, K.~F.}, \bibinfo{author}{McGill, K.~L.},
  \bibinfo{author}{Park, J.} \& \bibinfo{author}{McEuen, P.~L.}
\newblock \bibinfo{title}{The valley {H}all effect in {M}o{S}$_2$ transistors}.
\newblock \emph{\bibinfo{journal}{Science}} \textbf{\bibinfo{volume}{344}},
  \bibinfo{pages}{1489--1492} (\bibinfo{year}{2014}).

\bibitem{sodemann2015quantum}
\bibinfo{author}{Sodemann, I.} \& \bibinfo{author}{Fu, L.}
\newblock \bibinfo{title}{Quantum nonlinear {H}all effect induced by {B}erry
  curvature dipole in time-reversal invariant materials}.
\newblock \emph{\bibinfo{journal}{Phys. Rev. Lett.}}
  \textbf{\bibinfo{volume}{115}}, \bibinfo{pages}{216806}
  (\bibinfo{year}{2015}).

\bibitem{Iyama_2013}
\bibinfo{author}{Iyama, A.} \& \bibinfo{author}{Kimura, T.}
\newblock \bibinfo{title}{Magnetoelectric hysteresis loops in {C}r$_2${O}$_3$
  at room temperature}.
\newblock \emph{\bibinfo{journal}{Phys. Rev. B}} \textbf{\bibinfo{volume}{87}}
  (\bibinfo{year}{2013}).

\end{thebibliography}


\begin{thebibliography}{99}  
 \makeatletter
 \addtocounter{NAT@ctr}{58}
 \makeatother

\bibitem{zhao2019} Zhao, S. Y. F. \textit{et al.} Sign reversing Hall effect in atomically thin high temperature superconductors \textit{Phys. Rev. Lett.} $\mathbf{122}$, 166601 (2019).

\bibitem{deng2018} Deng, Y. \textit{et al.} Gate-tunable room-temperature ferromagnetism in two-dimensional Fe$_3$GeTe$_2$. \textit{Nature} $\mathbf{563}$, 94-99 (2018).

\bibitem{zhang2009} Deng, Y. \textit{et al.} Direct observation of a widely tunable bandgap in bilayer graphene. \textit{Nature} $\mathbf{459}$, 820-823 (2009).

\bibitem{taychatanapat2010} Taychatanapat, T. \& Jarillo-Herrero, P. Electronic transport in dual-gated bilayer graphene at large displacement fields. \textit{Phys. Rev. Lett.} $\mathbf{105}$, 166601 (2010).

\bibitem{VASP} Kresse, G. and Furthm\"uller, J. Efficient iterative schemes for \textit{ab initio} total-energy calculations using a plane-wave basis set. \textit{Phys. Rev. B} $\mathbf{54}$, 11169-11186 (1996).

\bibitem{wannier} Souza, I., Marzari, N. \&Vanderbilt, D. Maximally localized Wannier functions for entangled energy bands. \textit{Phys. Rev. B} $\mathbf{65}$, 035109 (2001).

\bibitem{Otrokov2019} Otrokov, M. M. \textit{et al.} Unique thickness-dependent properties of the van der Waals interlayer antiferromagnet MnBi$_2$Te$_4$ films. \textit{Phys. Rev. Lett.} $\mathbf{122}$, 107202 (2019).
\bibitem{yasuda2020} Yasuda, K. \textit{et al.} Stacking-engineered ferroelectricity in bilayer boron nitride. \textit{arXiv preprint} arXiv:2010.06600 (2020). 
\end{thebibliography}

\vspace{0.5cm}
\textbf{Methods}

\textbf{Crystal growth:} Our bulk crystals were grown by two methods: the Bi$_2$Te$_3$ flux method \cite{Yan2019} and solid-state reaction method with extra Mn and I$_2$. In the Bi$_2$Te$_3$ flux method, elemental Mn, Bi and Te were mixed at a molar ratio of $15:170:270$, loaded in a crucible, and sealed in a quartz tube under one-third atmospheric pressure of Ar. The ampule was first heated to $900^{\circ}$C for $5$ hours. It was then moved to another furnace where it slowly cooled from $597^{\circ}$C to $587^{\circ}$C and stayed for one day at $587^{\circ}$C. Finally, MnBi$_2$Te$_4$ were obtained by centrifuging the ampule to separate the crystals from the Bi$_2$Te$_3$ flux. In the solid-state reaction method, elemental form of Mn, Bi, Te and I$_2$ were first mixed at a stoichiometric ratio of $1.5:2:4:0.5$ and sealed in a quartz ampoule under vacuum. The sample was heated to $900^{\circ}$C in 24 hours in a box furnace and stayed at the temperature for over 5 hours to ensure a good mixture. The ampoule was then air quenched and moved to another furnace preheated at $597^{\circ}$C, where it then slowly cooled to $587^{\circ}$C in 72 hours and stayed at the final temperature for two weeks.

\textbf{Sample fabrication:} To address the sensitive chemical nature of 2D MnBi$_2$Te$_4$ flakes, all fabrication processes  were completed in an argon environment without exposure to air, chemicals, or heat. Specifically, the argon-filled glovebox with O$_2$ and H$_2$O level below $0.01$ ppm and a dew point below $-96^{\circ}$C was used. The glovebox was attached to an e-beam evaporator, allowing us to make metal deposition without exposure to air. First, thin flakes of MnBi$_2$Te$_4$ were mechanically exfoliated from the bulk crystal onto O$_2$ plasma cleaned $300$-nm SiO$_2/$Si wafers using Scotch-tape. Second, the number of layers were determined by the optical contrast of the flakes. This method has been proven effective for a wide range of air-sensitive vdW materials \cite{huang2017layer, deng2018, Deng2020} including MnBi$_2$Te$_4$. Specifically, the optical contrast ($C=\frac{I_{\textrm{flake}}-I_{\textrm{substrate}}}{I_{\textrm{flake}}+I_{\textrm{substrate}}}$) of MnBi$_2$Te$_4$ flakes with different thicknesses were first measured based on the optical images taken via a Nikon Eclipse LV150N microscope inside the glovebox (Extended Data Fig.~\ref{Extended_Data_Figure_Contrast}). By extracting and averaging individual RGB values over each flake and substrate region, we obtained the optical contrast ($C$). Then the flakes were taken out of the glovebox so that the number of layer was directly determined by atomic force microscopy. As such, a one-to-one correspondence between the optical contrast and the number of layer was established. This process was repeated many times to ensure that the correspondence was reproducible and reliable (see data from different samples in Extended Data Fig.~\ref{Extended_Data_Figure_Contrast}). For the flakes eventually selected to make devices, their thickness was determined by this optical contrast. After the transport measurements, their thickness was re-confirmed by atomic force microscopy or transmission electron microscopy. Third, a high-resolution stencil mask technique \cite{zhao2019, deng2018} was adapted. Specifically, the device contact pattern was written onto the exposed SiN membrane via photolithography and etched via reactive ion etching. This technique yields a resolution of $1$ $\mu$m. The mask was subsequently held in place using vacuum suction cups and aligned onto the MnBi$_2$Te$_4$ crystal by a microscope. Forth, the sample/mask assembly was transferred to an e-beam evaporator for metal evaporation without exposure to air. Fifth, a $20-50$ nm BN flake was transferred onto the MnBi$_2$Te$_4$ flake as the top gate dielectric. A graphite gate was transferred onto the MnBi$_2$Te$_4$/BN heterostructure.

\textbf{Electrical transport measurements with dual gating:} Electrical transport measurements were carried out in a PPMS (Quantum Design DynaCool). The base temperature was 1.65 K and maximum magnetic field was 9 T. The magnetic field was applied along the out-of-plane direction. Longitudinal and Hall voltages were measured simultaneously using standard lock-in techniques. The gate voltages were applied by Keithley 2400 source meters. In such dual-gated devices, as established by previous works \cite{zhang2009, taychatanapat2010}, the charge density $n$ and electric displacement field $D=\epsilon E$ can be controlled independently by the combination of top and bottom gate voltages $V_{\textrm{TG}}$ and $V_{\textrm{BG}}$. Specifically, the charge density can be obtained by $n=\frac{\epsilon_0\epsilon^{\textrm{hBN}}}{e}(V_{\textrm{TG}}-V_{\textrm{TG0}})/h_{\textrm{T}}+\frac{\epsilon_0\epsilon^{\textrm{SiO}_2}}{e}(V_{\textrm{BG}}-V_{\textrm{BG0}})/h_{\textrm{B}}$. The electric displacement field can be obtained by $D=[\epsilon^{\textrm{SiO}_2}(V_{\textrm{BG}}-V_{\textrm{BG0}})/h_{\textrm{B}}-\epsilon^{\textrm{hBN}}(V_{\textrm{TG}}-V_{\textrm{TG0}})/h_{\textrm{T}}]/2$. Here, $\epsilon_0=8.85\times10^{-12}$ F/m is the vacuum permittivity; $V_{\textrm{TG0}}$ and $V_{\textrm{BG0}}$ correspond to the gate voltages for the resistance maximum, which is the charge neutrality with no electric field; $\epsilon^{\textrm{hBN}} \sim 3$ and $\epsilon^{\textrm{SiO}_2}\sim 3.9$ are relative dielectric constants for hBN and SiO$_2$; , $h_{\textrm{T}}$ and $h_{\textrm{B}}$ are the thicknesses of the top hBN layer and bottom SiO$_2$ layers, respectively.

\textbf{First-principles calculations:} First-principles band structure calculations were performed using the projector augmented wave method as implemented in the VASP package \cite{VASP} within the generalized gradient approximation (GGA) schemes. 8$\times$8$\times$4 Monkhorst-Pack $k$-point meshes with an energy cutoff of $400$ eV were adapted for the Brillouin zone integration of bulk MnBi$_2$Te$_4$. Experimentally determined lattice parameters were used. In order to treat the localized Mn $3d$ orbitals, we follow previous first-principle works \cite{Otrokov2019a, Otrokov2019} on MnBi$_2$Te$_4$ and used an onsite $U~=~5.0$ eV. The Wannier model for the bulk structure was built using the Bi $p$ and Te $p$ orbitals \cite{wannier}. Thin films were modeled by constructing slabs of finite thickness using bulk Wannier model parameters.

\vspace{0.5cm}

\textbf{Data availability:} The data that support the plots within this paper and other findings of this study are available from the corresponding author upon reasonable request.

\textbf{Acknowledgement:} We gratefully thank Frank Zhao and Philip Kim for allowing us to use their glovebox and sample preparation facilities. We sincerely thank Trond I. Andersen, Giovanni Scuri, Hongkun Park and Mikhail D. Lukin for their help in magnetic measurements. We also thank Frank Zhao, Philip Kim, Yang Gao and Chang Liu for helpful discussions. Work in the Xu group was partly supported by the Center for the Advancement of Topological Semimetals (CATS), an Energy Frontier Research Center (EFRC) funded by the U.S. Department of Energy (DOE) Office of Science, through the Ames Laboratory under contract DE-AC0207CH11358 (fabrication and measurements) and partly through the STC Center for Integrated Quantum Materials (CIQM), National Science Foundation (NSF) award no. ECCS-2025158 (data analysis). SYX acknowledges the Corning Fund for Faculty Development. QM acknowledges support from the CATS, an EFRC funded by the U.S. DOE Office of Science, through the Ames Laboratory under contract DE-AC0207CH11358. CT acknowledges support from the Swiss National Science Foundation under project P2EZP2\_191801. YFL, AA (Akey), JG, DCB, and LF were supported by the STC Center for Integrated Quantum Materials (CIQM), NSF award no. ECCS-2025158. This work was performed in part at the Center for Nanoscale Systems (CNS) Harvard University, a member of the National Nanotechnology Coordinated Infrastructure Network (NNCI), which is supported by the NSF under NSF award no.1541959. Work at UCLA was supported by the U.S. DOE, office of Science, office of Basic Energy Sciences (BES) under Award Number DE-SC0021117 for bulk sample growth, transport and magnetic property measurements. Work at Northeastern University was supported by the US DOE, office of Science, office of BES grant number DE-SC0019275 and benefited from Northeastern University's Advanced Scientific Computation Center (ASCC) and the NERSC supercomputing center through DOE grant number DE-AC02-05CH11231. BG and AA (Agarwal) were supported by the Science Education and Research Board (SERB) and the Department of Science and Technology (DST) of the government of India for financial support, and the computer center IIT Kanpur, for providing the High Performance Computing facility. T-RC was supported by the Young Scholar Fellowship Program from the Ministry of Science and Technology (MOST) in Taiwan, under a MOST grant for the Columbus Program MOST110-2636-M-006-016, NCKU, Taiwan, and National Center for Theoretical Sciences, Taiwan. Work at NCKU was supported by the MOST, Taiwan, under grant MOST107-2627-E-006-001 and Higher Education Sprout Project, Ministry of Education to the Headquarters of University Advancement at NCKU. HL acknowledges the support by the MOST in Taiwan under grant number MOST 109-2112-M-001-014-MY3. HZL was supported by the National Natural Science Foundation of China (11925402), Guangdong province (2016ZT06D348, 2020KCXTD001), the National Key R \& D Program (2016YFA0301700), Shenzhen High-level Special Fund (G02206304, G02206404), and the Science, Technology and Innovation Commission of Shenzhen Municipality (ZDSYS20170303165926217, JCYJ20170412152620376, KYTDPT20181011104202253), and Center for Computational Science and Engineering of SUSTech. RC was supported by the China Postdoctoral Science Foundation (Grant No. 2019M661678) and the SUSTech Presidential Postdoctoral Fellowship. CF was supported by the ERC Advanced Grant No. 742068 `TOPMAT' and by the Deutsche Forschungsgemeinschaft (DFG, German Research Foundation) under Germany's Excellence Strategy through W\"urzburg-Dresden Cluster of Excellence on Complexity and Topology in Quantum Matter - ct.qmat (EXC 2147, project-id 390858490). KSB is grateful for the support of the Office of Naval Research under Award number N00014-20-1-2308. KW and TT acknowledge support from the Elemental Strategy Initiative conducted by the MEXT, Japan, Grant Number JPMXP0112101001 and JSPS KAKENHI Grant Number JP20H00354. ZZ, NW, ZH and WG thank Singapore National Research Foundation through its Competitive Research Program (CRP Award No. NRF-CRP21-2018-0007, NRF-CRP22-2019-0004). X-YZ, Y-XW, and BBZ acknowledge support from the NSF award No. ECCS-2041779.

\textbf{Author contributions:} SYX conceived the experiment and supervised the project. AG fabricated the devices with help from YFL, JXQ, DB, CF, KSB and QM. AG performed the transport measurements and analyzed data with help from YFL, CT, JXQ, SCH, DB, TD and QM. CH and NN grew the bulk MnBi$_2$Te$_4$ single crystals. ZZ, NW, ZH and WG as well as JXQ, CT and AG performed optical magnetic circular dichroism measurements. X-YZ, Y-XW, and BBZ performed nitrogen-vacancy center magnetometry experiments.  BG, RC, HS, AA (Agarwal), CT, SYX, H-ZL, H-JT, BS, AB, HL, LF, T-RC made theoretical studies including first-principles calculations and tight-binding modeling. AA (Akey), JG and DCB performed transmission electron microscope measurements. JXQ performed atomic force microscope measurements. KW and TT grew the bulk hBN single crystals. SYX, AG and QM wrote the manuscript with input from all authors. SYX was responsible for the overall direction, planning and integration among different research units.

\textbf{Competing financial interests:} The authors declare no competing financial interests.

\clearpage

\clearpage

\clearpage

\setcounter{figure}{0}
\renewcommand{\figurename}{\textbf{Extended Data Fig.}}

\clearpage
\begin{figure*}[t]
\includegraphics[width=15cm]{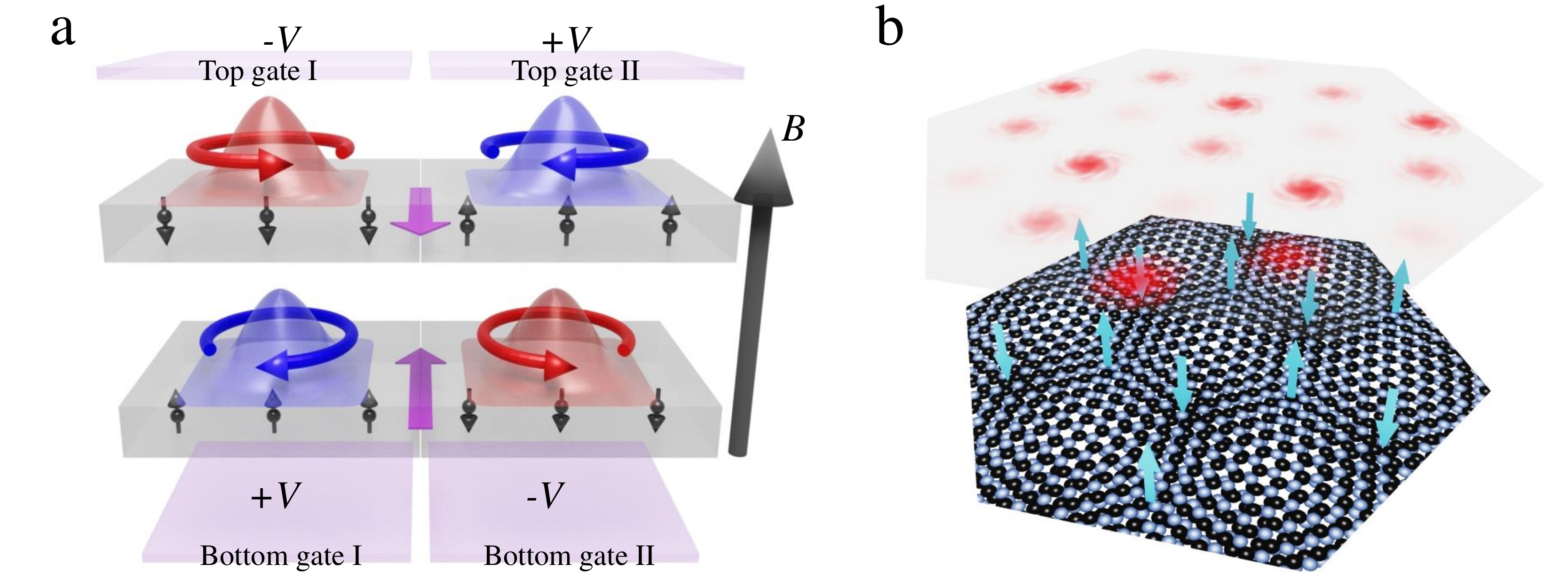}
\caption{Topological Axion domain wall constructed by the Axion field $\mathbf{E}\cdot\mathbf{B}$ (panel \textbf{a}) and spatially modulating Berry curvature moir\'e superlattice enabled by the MnBi$_2$Te$_4$-twisted hBN heterostructure \cite{yasuda2020} (panel \textbf{b}). }
\label{Extended_Data_Figure_Future}
\end{figure*}

\clearpage
\begin{figure*}[t]
\includegraphics[width=15cm]{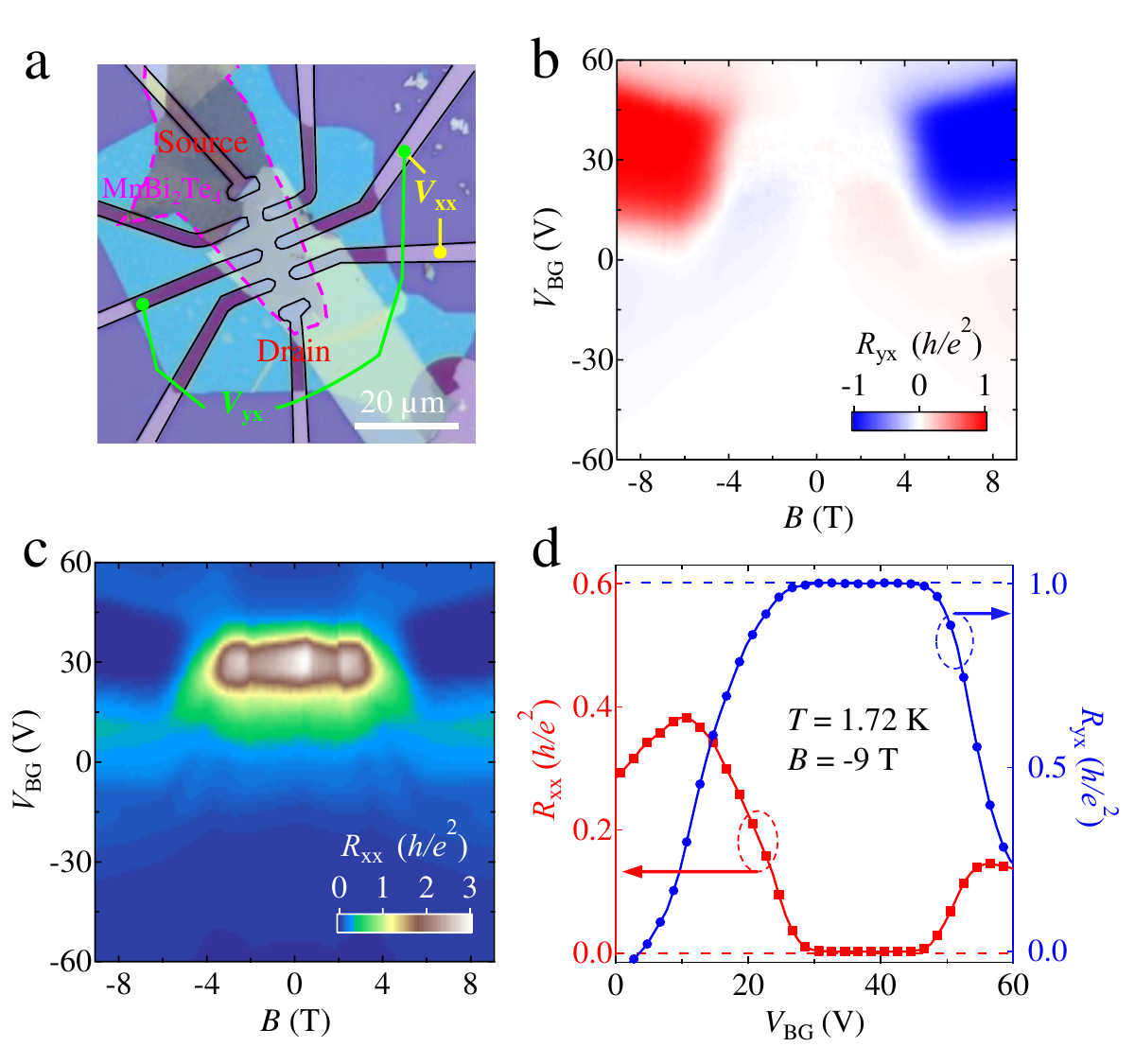}
\caption{\textbf{a,} Microscope image of our 6SL MnBi$_2$Te$_4$ device (the device presented in the main text). The circuit for our transport measurements is noted. \textbf{b,c,}  Longitudinal $R_{xx}$ (panel \textbf{c}) and transverse (Hall) resistance $R_{yx}$ (panel \textbf{b}) as a function of $V_{\textrm{BG}}$ and $B$. \textbf{d,} $R_{xx}$ and $R_{yx}$ $vs$ $V_{\textrm{BG}}$ at $-9$ T.}
\label{Extended_Data_Figure_QAHE}
\end{figure*}

\clearpage
\begin{figure*}[t]
\includegraphics[width=14cm]{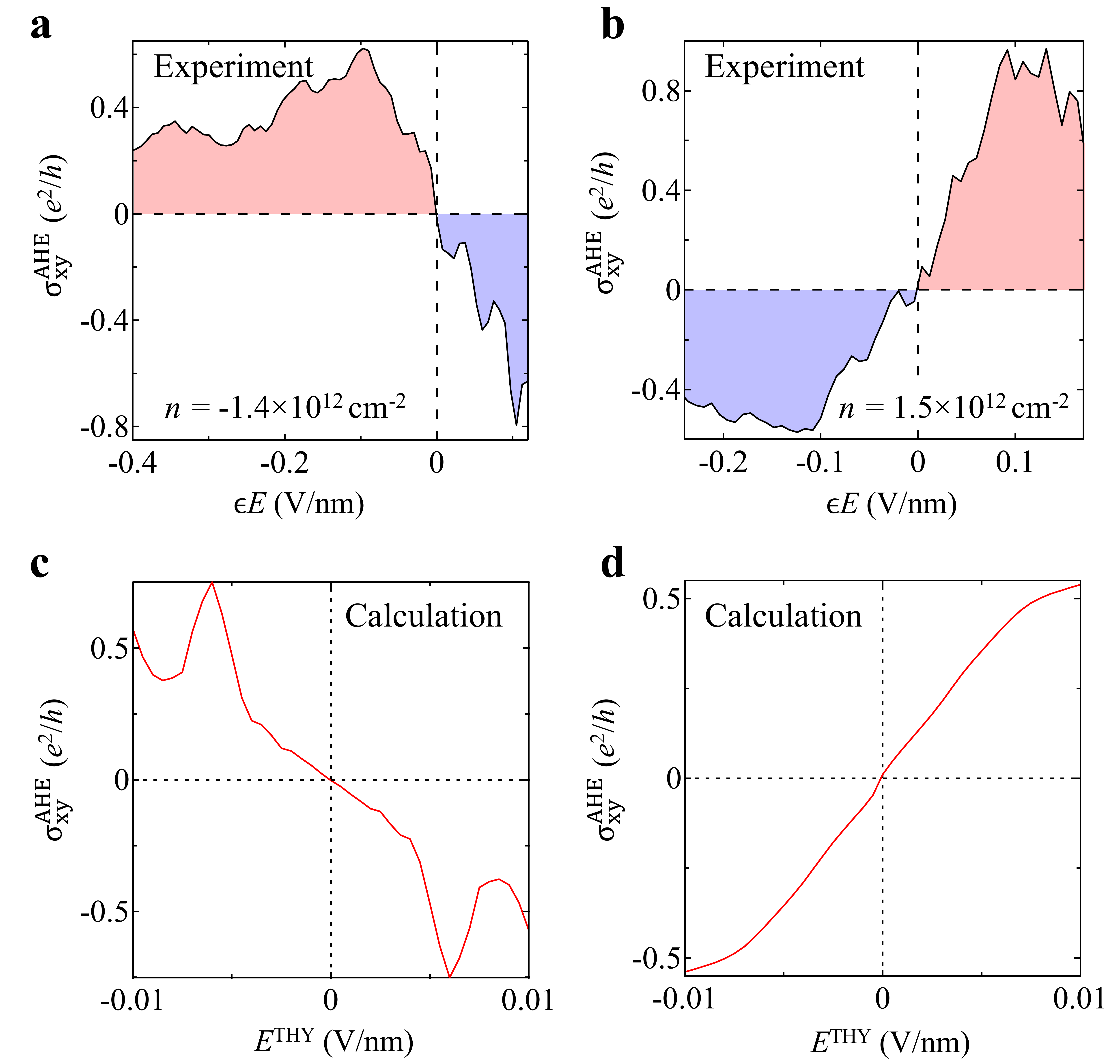}
\caption{\textbf{$E$ field dependence of the layer Hall effect in 6SL MnBi$_2$Te$_4$. a,} The AHE conductivity $\sigma_{xy}^{\textrm{AHE}}$ as a function of $E$ field. The charge density $n$ is set in the hole-doped regime ($n=-1.4\times10^{12}$ cm$^{-2}$). \textbf{b,} Same as panel (\textbf{a}) but $n$ is set in the electron-doped regime ($n=+1.5\times10^{12}$ cm$^{-2}$). \textbf{c,d,} First-principles calculated AHE conductivity $\sigma_{xy}^{\textrm{AHE}}$ as a function of $E$ field. \textbf{c,} Fermi level is set in the valence band ($-10$ meV). \textbf{d,} Fermi level is set in the conduction band ($+30$ meV).  }
\label{Extended_Data_Figure_Sigma_E_Calculation}
\end{figure*}

\begin{figure*}[h]
\includegraphics[width=14cm]{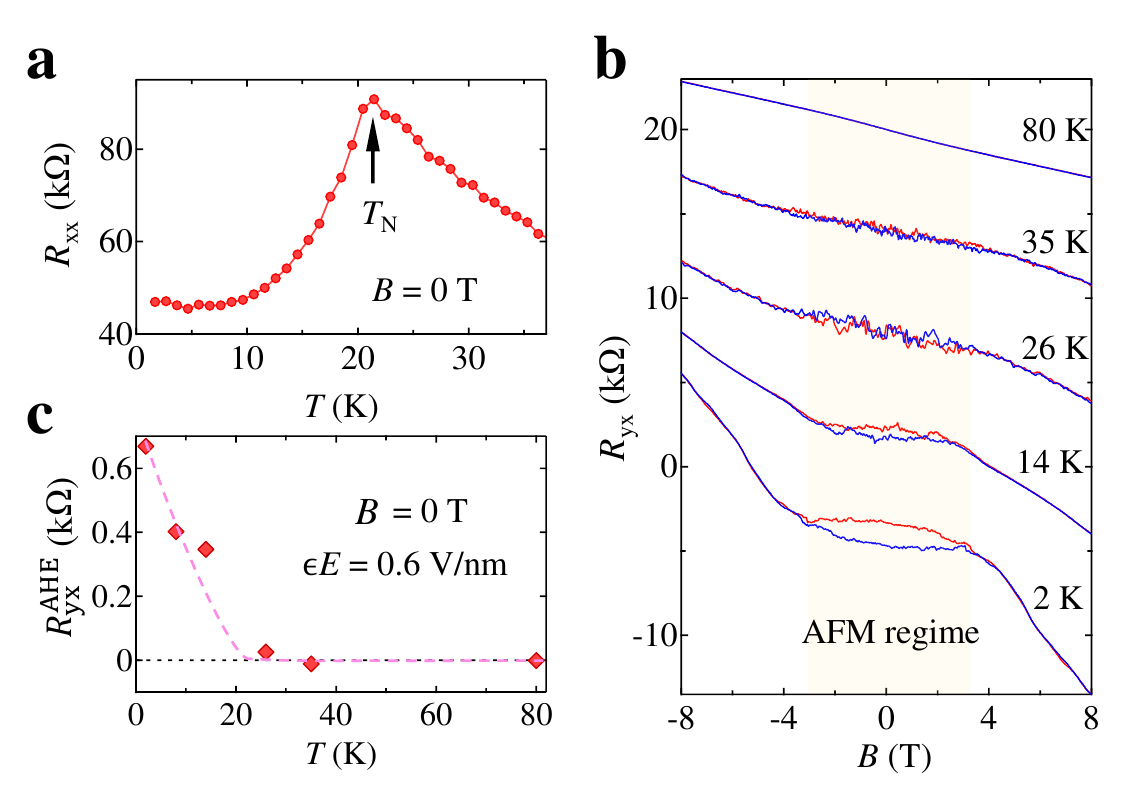}
\caption{{\bf Temperature dependent measurements of 6SL MnBi$_2$Te$_4$.} \textbf{a,} Temperature dependent $R_{xx}$ data showing the N\'eel temperature $T_{\textrm{N}}$. \textbf{b,} $R_{yx}-B$ measurements at different temperatures. Data at different temperatures are offset by $4$ kohm for visibility. \textbf{c,} AHE resistance as a function of temperature.}
\label{Extended_Data_Figure_Temperature_Dependence}
\end{figure*}

\begin{figure*}[h]
\includegraphics[width=16cm]{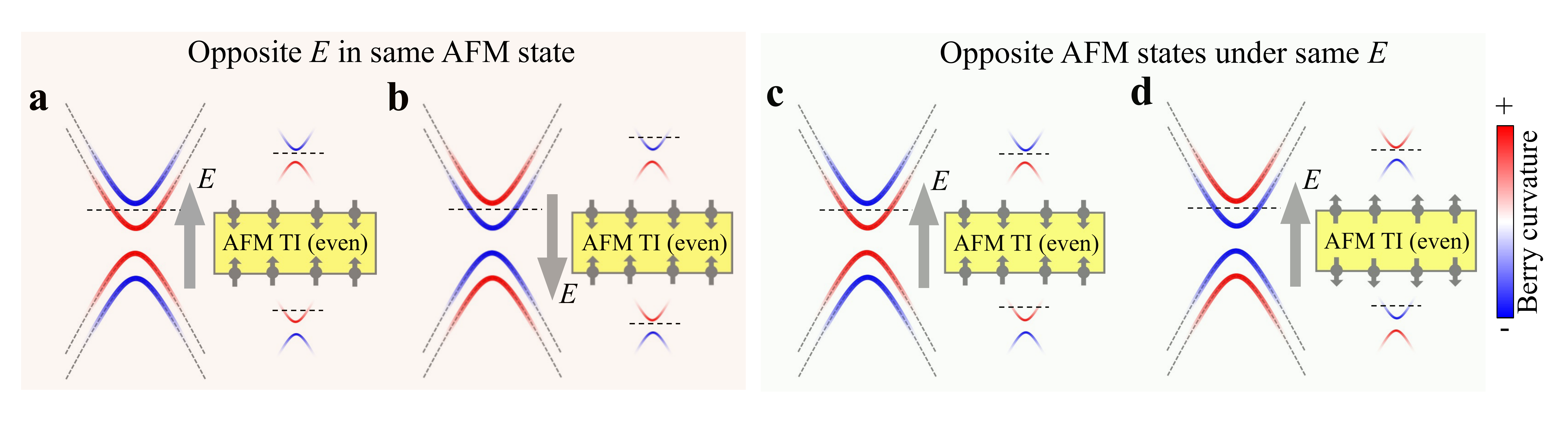}
\caption{Schematic electronic structure and Berry curvature of even-layered MnBi$_2$Te$_4$. \textbf{a,b,} Same AFM state under opposite $E$ field. \textbf{c,d,} opposite AFM states under same $E$ field.  }
\label{Opposite_Domain}
\end{figure*}

\begin{figure*}[h]
\includegraphics[width=17cm]{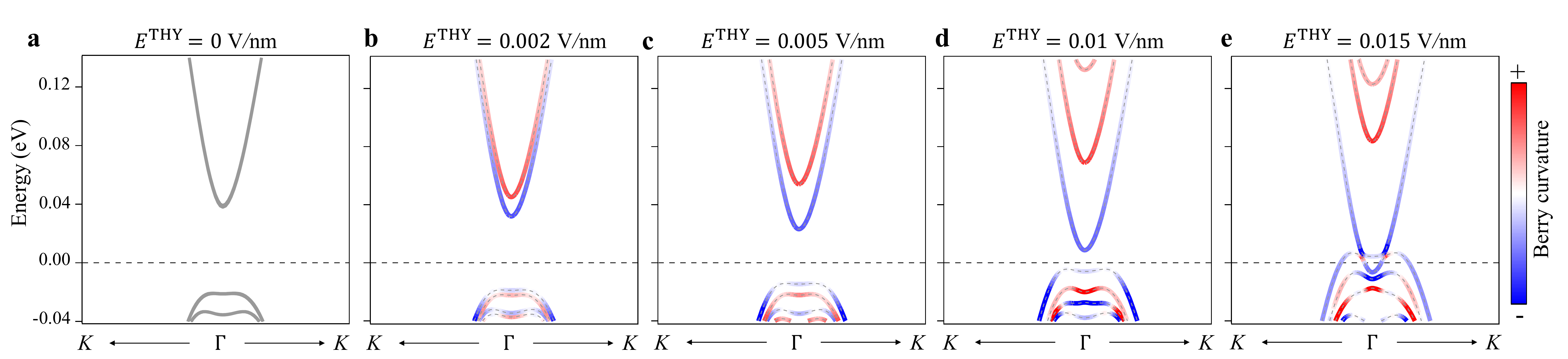}
\caption{{\bf First-principles calculated band structures as a function of electric field.}}
\label{Extended_Data_Figure_DFT}
\end{figure*}

\begin{figure*}[h]
\includegraphics[width=13cm]{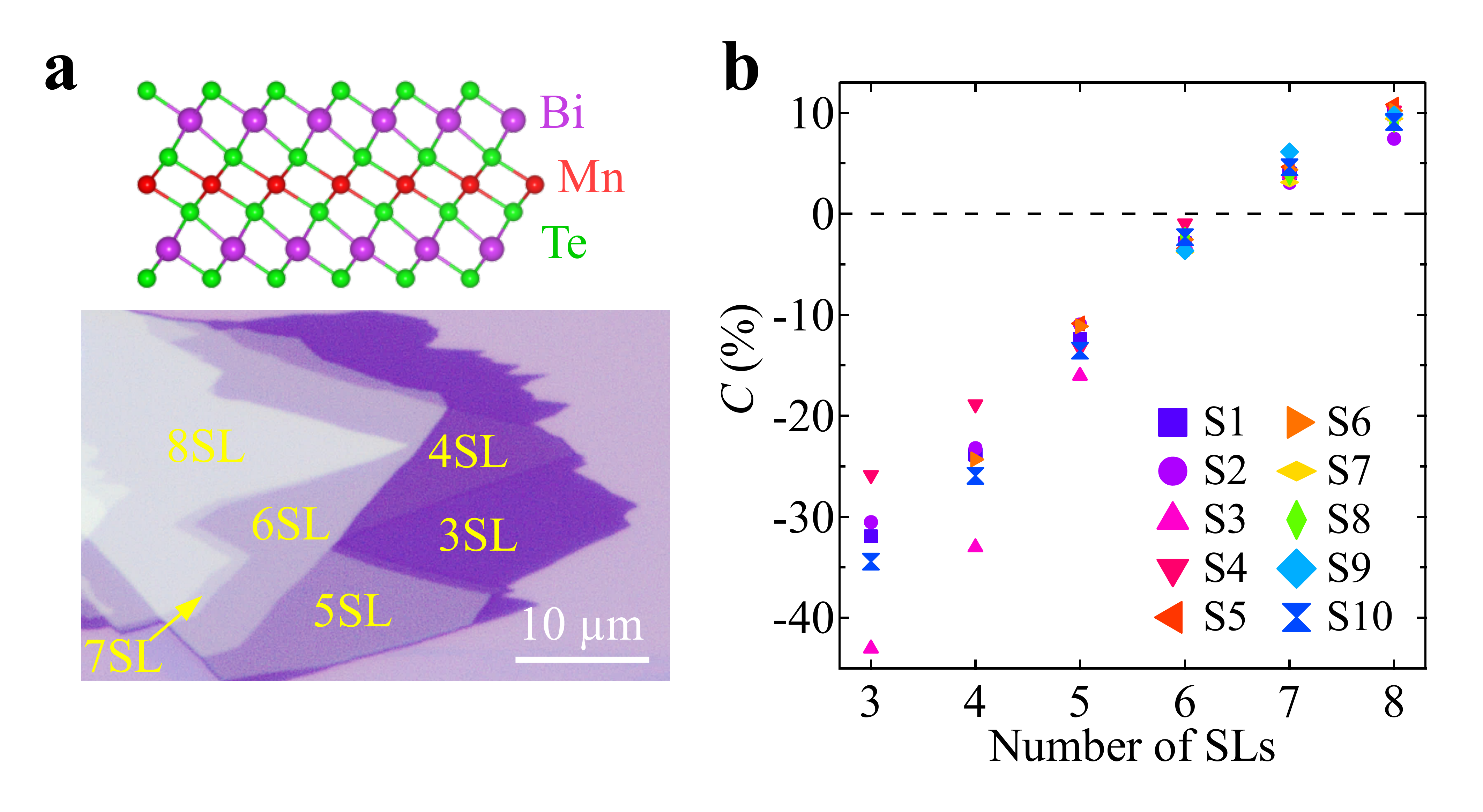}
\caption{{\bf a,} Lower: Optical image of few-layer flakes of MnBi$_2$Te$_4$ exfoliated on SiO$_2$ substrate. Upper: Lattice of one SL MnBi$_2$Te$_4$. \textbf{b,} Optical contrast ($C=\frac{I_{\textrm{flake}}-I_{\textrm{substrate}}}{I_{\textrm{flake}}+I_{\textrm{substrate}}}$) as a function of number of layers, which was independently determined by atomic force microscope. This process was repeated on many samples (each symbol in the figure represent an independent sample) to ensure a reproducible and reliable correspondence between $C$ and layer number. }
\label{Extended_Data_Figure_Contrast}
\end{figure*}

\begin{figure*}[h]
\includegraphics[width=17cm]{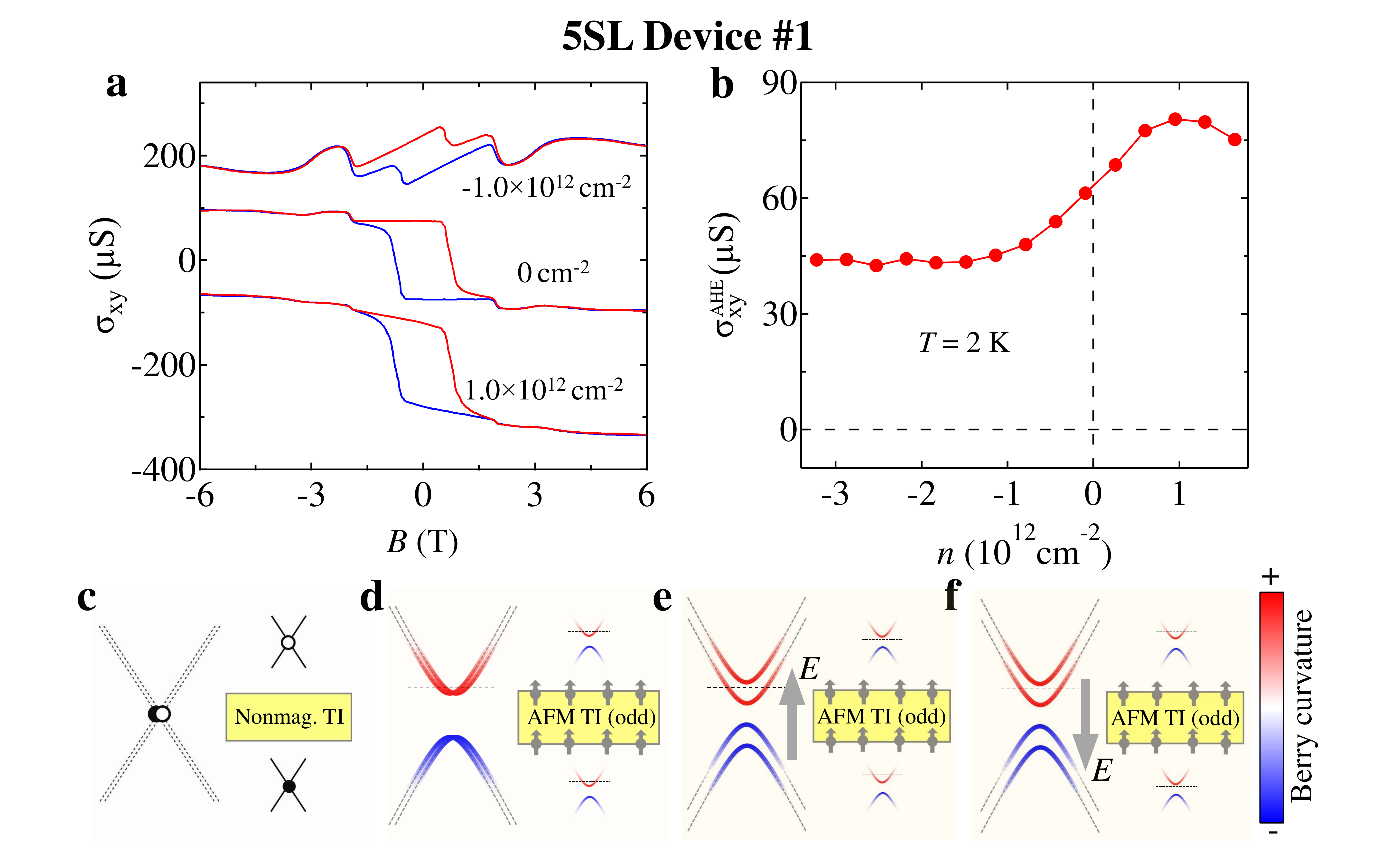}
\caption{{\bf Experimental data and microscopic picture for odd-layered MnBi$_2$Te$_4$.} \textbf{a,b} 5SL MnBi$_2$Te$_4$'s AHE. In contrast to 6SL, the AHE in 5SL does not change sign as one tunes the charge density from hole-doped regime to electron-doped regime. Data at different $n$ are offset by $200$ $\mu$S for visibility. \textbf{c-f}. In an odd-layered AFM system, the top and bottom Dirac fermions experience same magnetizations and hence open up gaps in the same fashion. As such, conduction and valence bands have the opposite Berry curvature. Therefore, the AHE remains the same sign in the hole-doped and electron-doped regimes. This conclusion is independent of $E$.}
\label{Extended_Data_Figure_5SL}

\end{figure*}

\end{document}